\documentclass[twocolumn,letterpaper,aps,prc,longbibliography,superscriptaddress,nofootinbib,floatfix]{revtex4-2}

\usepackage{graphicx}   % Include figure files
\usepackage{xspace}

\newcommand{\pt}{\mbox{$p_T$}\xspace}

\newcommand{\raa}{\mbox{$R_{AA}$}\xspace}
\newcommand{\Npart}{\mbox{$\langle N_{\rm part}\rangle$}\xspace}
\newcommand{\Ncoll}{\mbox{$\langle N_{\rm coll} \rangle$}\xspace}

\newcommand{\sqs}{\mbox{$\sqrt{s}$}\xspace}
\newcommand{\sqstwo}{\mbox{$\sqrt{s}=200$~GeV}\xspace}
\newcommand{\sqsntwo}{\mbox{$\sqrt{s_{_{NN}}}=200$~GeV}\xspace}

\newcommand{\pp}{\mbox{$p$$+$$p$}\xspace}

\newcommand{\auau}{\mbox{Au$+$Au}\xspace}

\newcommand{\cuau}{\mbox{Cu$+$Au}\xspace}

\newcommand{\pbpb}{\mbox{Pb$+$Pb}\xspace}
\newcommand{\accEff}{$A\varepsilon$\xspace}

\newcommand{\Jpsi}{\mbox{$J/\psi$}\xspace}

\newcommand{\omgRho}{\mbox{$\omega+\rho$}\xspace}

\begin{document}

\title{Low-mass vector-meson production at forward rapidity in $p$$+$$p$ and 
Au$+$Au collisions at $\sqrt{s_{_{NN}}}=200$~GeV}

\date{\today}

\newcommand{\abilene}{Abilene Christian University, Abilene, Texas 79699, USA}
\newcommand{\augie}{Department of Physics, Augustana University, Sioux Falls, South Dakota 57197, USA}
\newcommand{\banaras}{Department of Physics, Banaras Hindu University, Varanasi 221005, India}
\newcommand{\barc}{Bhabha Atomic Research Centre, Bombay 400 085, India}
\newcommand{\baruch}{Baruch College, City University of New York, New York, New York, 10010 USA}
\newcommand{\bnlcoll}{Collider-Accelerator Department, Brookhaven National Laboratory, Upton, New York 11973-5000, USA}
\newcommand{\bnlphys}{Physics Department, Brookhaven National Laboratory, Upton, New York 11973-5000, USA}
\newcommand{\caucr}{University of California-Riverside, Riverside, California 92521, USA}
\newcommand{\charlesczech}{Charles University, Faculty of Mathematics and Physics, 180 00 Troja, Prague, Czech Republic}
\newcommand{\ciae}{Science and Technology on Nuclear Data Laboratory, China Institute of Atomic Energy, Beijing 102413, People's Republic of China}
\newcommand{\cns}{Center for Nuclear Study, Graduate School of Science, University of Tokyo, 7-3-1 Hongo, Bunkyo, Tokyo 113-0033, Japan}
\newcommand{\colorado}{University of Colorado, Boulder, Colorado 80309, USA}
\newcommand{\columbia}{Columbia University, New York, New York 10027 and Nevis Laboratories, Irvington, New York 10533, USA}
\newcommand{\czechtech}{Czech Technical University, Zikova 4, 166 36 Prague 6, Czech Republic}
\newcommand{\debrecen}{Debrecen University, H-4010 Debrecen, Egyetem t{\'e}r 1, Hungary}
\newcommand{\elte}{ELTE, E{\"o}tv{\"o}s Lor{\'a}nd University, H-1117 Budapest, P{\'a}zm{\'a}ny P.~s.~1/A, Hungary}
\newcommand{\ewha}{Ewha Womans University, Seoul 120-750, Korea}
\newcommand{\fsu}{Florida State University, Tallahassee, Florida 32306, USA}
\newcommand{\gsu}{Georgia State University, Atlanta, Georgia 30303, USA}
\newcommand{\hiroshima}{Physics Program and International Institute for Sustainability with Knotted Chiral Meta Matter (WPI-SKCM2), Hiroshima University, Higashi-Hiroshima, Hiroshima 739-8526, Japan}
\newcommand{\howard}{Department of Physics and Astronomy, Howard University, Washington, DC 20059, USA}
\newcommand{\hunrenatomki}{HUN-REN ATOMKI, H-4026 Debrecen, Bem t{\'e}r 18/c, Hungary}
\newcommand{\ihepprot}{IHEP Protvino, State Research Center of Russian Federation, Institute for High Energy Physics, Protvino, 142281, Russia}
\newcommand{\illuiuc}{University of Illinois at Urbana-Champaign, Urbana, Illinois 61801, USA}
\newcommand{\inrras}{Institute for Nuclear Research of the Russian Academy of Sciences, prospekt 60-letiya Oktyabrya 7a, Moscow 117312, Russia}
\newcommand{\instpasczech}{Institute of Physics, Academy of Sciences of the Czech Republic, Na Slovance 2, 182 21 Prague 8, Czech Republic}
\newcommand{\isu}{Iowa State University, Ames, Iowa 50011, USA}
\newcommand{\jaea}{Advanced Science Research Center, Japan Atomic Energy Agency, 2-4 Shirakata Shirane, Tokai-mura, Naka-gun, Ibaraki-ken 319-1195, Japan}
\newcommand{\jeonbuk}{Jeonbuk National University, Jeonju, 54896, Korea}
\newcommand{\jyvaskyla}{Helsinki Institute of Physics and University of Jyv{\"a}skyl{\"a}, P.O.Box 35, FI-40014 Jyv{\"a}skyl{\"a}, Finland}
\newcommand{\kek}{KEK, High Energy Accelerator Research Organization, Tsukuba, Ibaraki 305-0801, Japan}
\newcommand{\korea}{Korea University, Seoul 02841, Korea}
\newcommand{\kurchatov}{National Research Center ``Kurchatov Institute", Moscow, 123098 Russia}
\newcommand{\kyoto}{Kyoto University, Kyoto 606-8502, Japan}
\newcommand{\lawllnl}{Lawrence Livermore National Laboratory, Livermore, California 94550, USA}
\newcommand{\losalamos}{Los Alamos National Laboratory, Los Alamos, New Mexico 87545, USA}
\newcommand{\lund}{Department of Physics, Lund University, Box 118, SE-221 00 Lund, Sweden}
\newcommand{\lyon}{IPNL, CNRS/IN2P3, Univ Lyon, Universit{\'e} Lyon 1, F-69622, Villeurbanne, France}
\newcommand{\maryland}{University of Maryland, College Park, Maryland 20742, USA}
\newcommand{\mass}{Department of Physics, University of Massachusetts, Amherst, Massachusetts 01003-9337, USA}
\newcommand{\mate}{MATE, Institute of Technology, Laboratory of Femtoscopy, K\'aroly R\'obert Campus, H-3200 Gy\"ongy\"os, M\'atrai \'ut 36, Hungary}
\newcommand{\michigan}{Department of Physics, University of Michigan, Ann Arbor, Michigan 48109-1040, USA}
\newcommand{\miss}{Mississippi State University, Mississippi State, Mississippi 39762, USA}
\newcommand{\muhlenberg}{Muhlenberg College, Allentown, Pennsylvania 18104-5586, USA}
\newcommand{\nara}{Nara Women's University, Kita-uoya Nishi-machi Nara 630-8506, Japan}
\newcommand{\natmephi}{National Research Nuclear University, MEPhI, Moscow Engineering Physics Institute, Moscow, 115409, Russia}
\newcommand{\newmex}{University of New Mexico, Albuquerque, New Mexico 87131, USA}
\newcommand{\nmsu}{New Mexico State University, Las Cruces, New Mexico 88003, USA}
\newcommand{\northcg}{Physics and Astronomy Department, University of North Carolina at Greensboro, Greensboro, North Carolina 27412, USA}
\newcommand{\ohio}{Department of Physics and Astronomy, Ohio University, Athens, Ohio 45701, USA}
\newcommand{\ornl}{Oak Ridge National Laboratory, Oak Ridge, Tennessee 37831, USA}
\newcommand{\orsay}{IPN-Orsay, Univ.~Paris-Sud, CNRS/IN2P3, Universit\'e Paris-Saclay, BP1, F-91406, Orsay, France}
\newcommand{\peking}{Peking University, Beijing 100871, People's Republic of China}
\newcommand{\pnpi}{PNPI, Petersburg Nuclear Physics Institute, Gatchina, Leningrad region, 188300, Russia}
\newcommand{\riken}{RIKEN Nishina Center for Accelerator-Based Science, Wako, Saitama 351-0198, Japan}
\newcommand{\rikjrbrc}{RIKEN BNL Research Center, Brookhaven National Laboratory, Upton, New York 11973-5000, USA}
\newcommand{\rikkyo}{Physics Department, Rikkyo University, 3-34-1 Nishi-Ikebukuro, Toshima, Tokyo 171-8501, Japan}
\newcommand{\saispbstu}{Saint Petersburg State Polytechnic University, St.~Petersburg, 195251 Russia}
\newcommand{\seoulnat}{Department of Physics and Astronomy, Seoul National University, Seoul 151-742, Korea}
\newcommand{\stonybrkc}{Chemistry Department, Stony Brook University, SUNY, Stony Brook, New York 11794-3400, USA}
\newcommand{\stonycrkp}{Department of Physics and Astronomy, Stony Brook University, SUNY, Stony Brook, New York 11794-3800, USA}
\newcommand{\tenn}{University of Tennessee, Knoxville, Tennessee 37996, USA}
\newcommand{\titech}{Department of Physics, Tokyo Institute of Technology, Oh-okayama, Meguro, Tokyo 152-8551, Japan}
\newcommand{\tsukuba}{Tomonaga Center for the History of the Universe, University of Tsukuba, Tsukuba, Ibaraki 305, Japan}
\newcommand{\usmma}{United States Merchant Marine Academy, Kings Point, New York 11024, USA}
\newcommand{\vandy}{Vanderbilt University, Nashville, Tennessee 37235, USA}
\newcommand{\weizmann}{Weizmann Institute, Rehovot 76100, Israel}
\newcommand{\wigner}{Institute for Particle and Nuclear Physics, HUN-REN Wigner Research Centre for Physics, (HUN-REN Wigner RCP, RMI), H-1525 Budapest 114, POBox 49, Budapest, Hungary}
\newcommand{\yonsei}{Yonsei University, IPAP, Seoul 120-749, Korea}
\newcommand{\zagreb}{Department of Physics, Faculty of Science, University of Zagreb, Bijeni\v{c}ka c.~32 HR-10002 Zagreb, Croatia}
\newcommand{\zambia}{Department of Physics, School of Natural Sciences, University of Zambia, Great East Road Campus, Box 32379, Lusaka, Zambia}
\affiliation{\abilene}
\affiliation{\augie}
\affiliation{\banaras}
\affiliation{\barc}
\affiliation{\baruch}
\affiliation{\bnlcoll}
\affiliation{\bnlphys}
\affiliation{\caucr}
\affiliation{\charlesczech}
\affiliation{\ciae}
\affiliation{\cns}
\affiliation{\colorado}
\affiliation{\columbia}
\affiliation{\czechtech}
\affiliation{\debrecen}
\affiliation{\elte}
\affiliation{\ewha}
\affiliation{\fsu}
\affiliation{\gsu}
\affiliation{\hiroshima}
\affiliation{\howard}
\affiliation{\hunrenatomki}
\affiliation{\ihepprot}
\affiliation{\illuiuc}
\affiliation{\inrras}
\affiliation{\instpasczech}
\affiliation{\isu}
\affiliation{\jaea}
\affiliation{\jeonbuk}
\affiliation{\jyvaskyla}
\affiliation{\kek}
\affiliation{\korea}
\affiliation{\kurchatov}
\affiliation{\kyoto}
\affiliation{\lawllnl}
\affiliation{\losalamos}
\affiliation{\lund}
\affiliation{\lyon}
\affiliation{\maryland}
\affiliation{\mass}
\affiliation{\mate}
\affiliation{\michigan}
\affiliation{\miss}
\affiliation{\muhlenberg}
\affiliation{\nara}
\affiliation{\natmephi}
\affiliation{\newmex}
\affiliation{\nmsu}
\affiliation{\northcg}
\affiliation{\ohio}
\affiliation{\ornl}
\affiliation{\orsay}
\affiliation{\peking}
\affiliation{\pnpi}
\affiliation{\riken}
\affiliation{\rikjrbrc}
\affiliation{\rikkyo}
\affiliation{\saispbstu}
\affiliation{\seoulnat}
\affiliation{\stonybrkc}
\affiliation{\stonycrkp}
\affiliation{\tenn}
\affiliation{\titech}
\affiliation{\tsukuba}
\affiliation{\usmma}
\affiliation{\vandy}
\affiliation{\weizmann}
\affiliation{\wigner}
\affiliation{\yonsei}
\affiliation{\zagreb}
\affiliation{\zambia}
\author{N.J.~Abdulameer} \affiliation{\debrecen} \affiliation{\hunrenatomki}
\author{U.~Acharya} \affiliation{\gsu}
\author{A.~Adare} \affiliation{\colorado} 
\author{C.~Aidala} \affiliation{\michigan} 
\author{N.N.~Ajitanand} \altaffiliation{Deceased} \affiliation{\stonybrkc} 
\author{Y.~Akiba} \email[PHENIX Spokesperson: ]{akiba@rcf.rhic.bnl.gov} \affiliation{\riken} \affiliation{\rikjrbrc}
\author{M.~Alfred} \affiliation{\howard} 
\author{D.~Anderson} \affiliation{\isu}
\author{V.~Andrieux} \affiliation{\michigan} 
\author{S.~Antsupov} \affiliation{\saispbstu}
% ...
\collaboration{PHENIX Collaboration}  \noaffiliation

%------------------------------------------------------------------------------|

\begin{abstract}

%\linenumbers

The PHENIX experiment at the Relativistic Heavy Ion Collider has 
measured low-mass vector-meson ($\omega+\rho$ and $\phi$) production 
through the dimuon decay channel at forward rapidity 
$(1.2<|\mbox{y}|<2.2)$ in $p$$+$$p$ and Au$+$Au collisions at 
$\sqrt{s_{_{NN}}}=200$~GeV. The low-mass vector-meson yield and 
nuclear-modification factor were measured as a function of the average 
number of participating nucleons, $\langle N_{\rm part}\rangle$, and the 
transverse momentum $p_T$. These results were compared with those 
obtained via the kaon decay channel in a similar $p_T$ range at 
midrapidity. The nuclear-modification factors in both rapidity regions 
are consistent within the uncertainties. A comparison of the 
$\omega+\rho$ and $J/\psi$ mesons reveals that the light and heavy 
flavors are consistently suppressed across both $p_T$ and 
${\langle}N_{\rm part}\rangle$. In contrast, the $\phi$ meson displays a 
nuclear-modification factor consistent with unity, suggesting 
strangeness enhancement in the medium formed.

\end{abstract}

\maketitle

\section{INTRODUCTION}

The formation of quark-gluon plasma (QGP) has been confirmed through 
experiments at both the Relativistic Heavy Ion Collider 
(RHIC)~\cite{Nucl.Phys.A757.184,Nucl.Phys.A757.1,Nucl.Phys.A757.28,Nucl.Phys.A757.102} 
and the Large Hadron Collider 
(LHC)~\cite{Eur.Phys.J.C72.1945,Phys.Lett.B720.52,Phys.Lett.B719.220,Nucl.Phys.A956.11}. 
After the initial confirmation, one of the primary objectives in 
high-energy nuclear physics, including the PHENIX experiment, has been 
to measure and explore the properties of the QGP. Strangeness, in 
particular, serves as an excellent probe of the QGP because strange 
quarks are produced in the early stages of the collision, and their 
behavior provides valuable insights into the QGP properties and 
dynamics.

The QGP is expected to enhance the production of strange quarks due to 
the relatively large mass of the strange quark and the abundance of 
gluons at high 
energies~\cite{KOCH1986167,PhysRevLett.54.1122,ALESSANDRO2003147,PhysRevC.72.014903,PhysRevLett.99.052301,PhysRevLett.99.112301,ADAMOVA2008425,PhysRevC.78.044907,ABELEV2009183,PhysRevC.83.024909,ARNALDI2011325}. 
Studying the strange quark production and its interactions in heavy-ion 
collisions aids in the exploration of the QGP properties and the 
mechanisms governing quark and gluon interactions within it. Strange 
hadrons, such as hyperons and strange mesons, are sensitive to the 
temperature and chemical potential of the QGP. Their yields and momentum 
distributions can be used to extract information about the temperature 
and chemical potential of the QGP, providing essential thermodynamic 
data~\cite{PhysRevLett.48.1066,rennecke2019strangenessneutralityqcdphase}.

The $\phi$ meson, with its nearly pure bound state of strange ($s$) and 
anti-strange ($\bar{s}$) quarks, provides an ideal system for studying 
strangeness enhancement. It is the lightest nearly pure bound state of 
strange quarks ($s\bar{s}$) with a mass comparable to the mass of the 
lightest baryons, such as the proton, and can be measured at PHENIX for 
$p_T>2.5$ GeV/$c$~over a broad rapidity range. Due to its longer lifetime 
(42~fm/$c$)~\cite{PhysRevLett.54.1122}, which is much longer than the QGP 
lifetime (5~fm/$c$)~\cite{Nucl.Phys.A757.184}, the $\phi$ meson is less 
influenced by late hadronic rescattering, thus reflecting the initial 
evolution of the system. As a pure $s\bar{s}$ state, the $\phi$ meson 
also places constraints on models describing parton energy loss and 
recombination in the QGP~\cite{EurPhysJC.77.571,PhysRevC.106.014908}. 
Therefore, the properties of the $\phi$ meson are primarily governed by 
conditions in the early partonic phase, making it an ideal probe to 
investigate the properties of matter created in relativistic ion 
collisions.

The production at RHIC of low-mass vector mesons (LVMs), such as the 
$\phi$, $\omega$, and $\rho$ mesons, offers important insights into QGP, 
especially when studied through their dimuon decay channel. Leptons are 
not affected by the strong color field within the QGP and can travel to 
the detectors with minimal interference. As a result, studying LVM 
production via their dimuon decay channel is an excellent way to explore 
evolution of the system. The $\omega$ meson, composed of light valence 
quarks similar to the $\pi^0$ but with a larger mass (782 
MeV/$c^2$)~\cite{J.Phys.G33.1,WANG2004299}, serves as an additional tool 
for systematically examining parton energy loss and hadron production 
mechanisms in these collisions. While the measurement of the $\rho$ 
meson spectral function can provide insights into in-medium 
modifications of hadron properties near the quantum-chromodynamics phase 
boundary, which are related to chiral symmetry 
restoration~\cite{VANHEES2008339, PhysRevD.86.034030, Petreczky_2012}, 
resolving the $\rho$ meson (770 MeV/$c^2$)~\cite{J.Phys.G33.1} spectral 
function in the two-muon channel requires better mass resolution than 
the muon spectrometers of the PHENIX experiment can offer. Due to this 
limitation, the $\rho$ meson cannot be distinguished from the $\omega$ 
meson, so the combined measurement of $\omega + \rho$ is typically 
reported.

Furthermore, measurements of the \omgRho and $\phi$ mesons, when 
combined with earlier \Jpsi data~\cite{PhysRevLett.98.232301} in the 
same kinematic region, offer a valuable opportunity to investigate the 
flavor dependence of medium effects. The $\omega + \rho$, $\phi$, and 
\Jpsi mesons comprise a closed system of light, strange, and charm 
quarks, respectively, and comparing the production of these mesons could 
provide insight into their respective interactions and behavior in the 
medium.

In this paper, the production of LVMs, \omgRho and $\phi$ mesons, is 
determined, in \pp and \auau collisions at \sqsntwo in the forward and 
backward rapidity regions, using dimuons detected by the PHENIX muon 
spectrometers. The invariant yields of \omgRho and $\phi$ mesons in \pp 
collisions will serve as a baseline for the \auau measurements. Although 
the production in \pp collisions was measured previously by 
PHENIX~\cite{PhysRevD.90.052002}, changes in the detector setup between 
the \auau and \pp data collections necessitate extracting the \pp 
invariant yield from a more recent data set to account for any potential 
variations in systematic effects. Measurements of the 
LVM nuclear-modification factor (\raa) as a function of \pt and the 
average number of participating nucleons are also presented. The results 
discussed are based on \auau collisions at \sqsntwo recorded in 2014 
with an integrated luminosity of 7.5 nb$^{-1}$, and \pp collisions at 
\sqstwo recorded in 2015 with an integrated luminosity of 47 pb$^{-1}$.

%%%%%%%%%%%%%%%%%%%%%%%%%%%%%%%%%%%%%%%%%%%%%%%%%%%
\section{Experimental Setup}
\label{sec:apparatus}

%%%%%%%%%%%%%%%%%%%%%%%%%%%%%%%%%%%%%%%%%%%%%%%%%%%%%%%%%%%%%%%%% Fig_1
\begin{figure}[htb]
\centering
\includegraphics[width=1.0\linewidth]{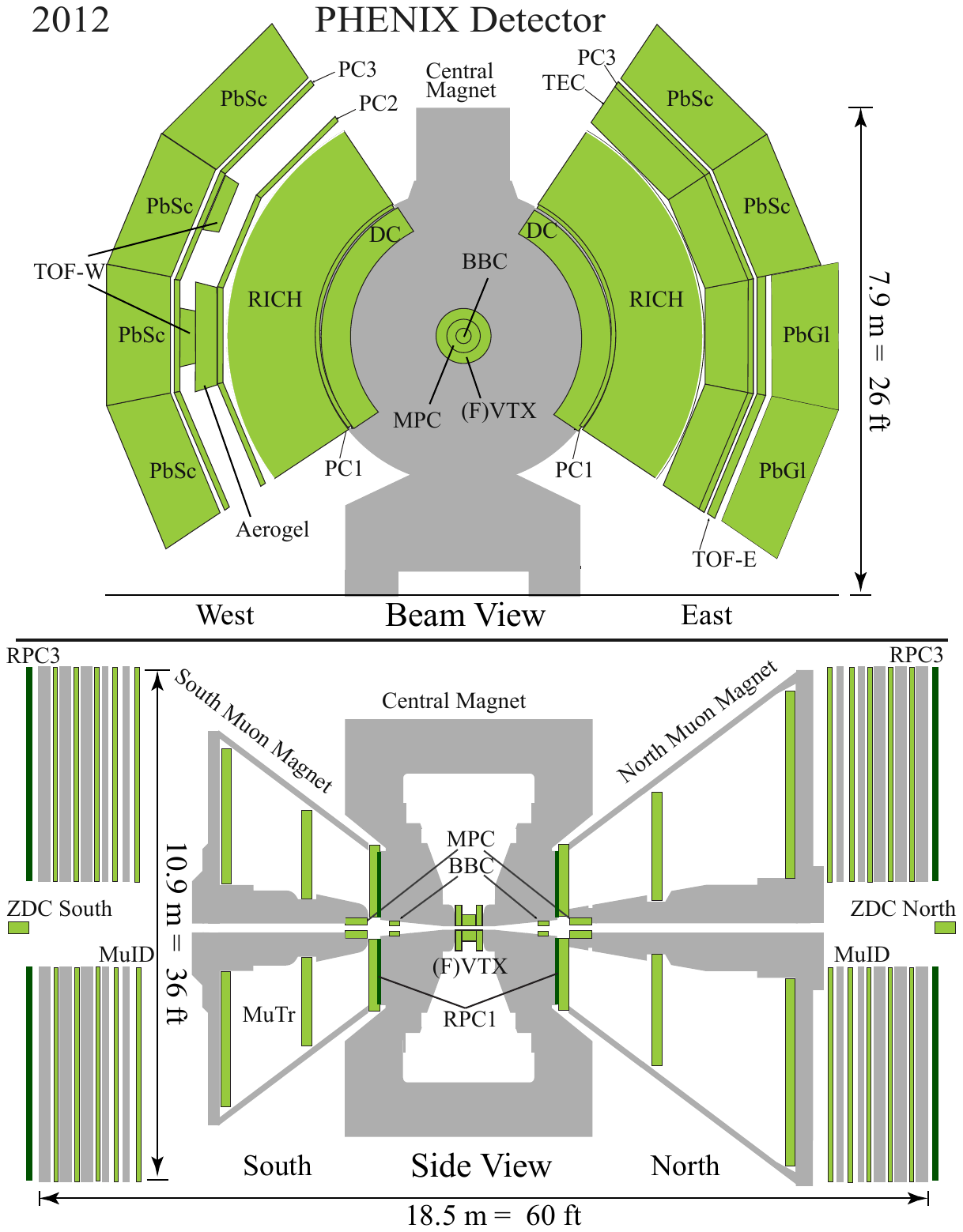}
\caption{\label{fig:Detector} 
A side view of the PHENIX detector, concentrating on the muon spectrometers 
instrumentation.}
\end{figure}

A detailed description of the PHENIX detector is available in 
Ref.~\cite{ADCOX2003469}. Here, we provide a brief overview of the 
detector subsystems relevant to these measurements. The key systems, 
depicted in Fig.~\ref{fig:Detector}, comprise forward silicon-vertex 
trackers (FVTX)~\cite{AIDALA201444}, hadron absorbers, and muon 
spectrometers~\cite{AKIKAWA2003537}.

The FVTX is a silicon detector designed to accurately measure the 
collision vertex, with additional data provided by the silicon vertex 
tracker (VTX) at midrapidity. Together, the VTX and FVTX determine the 
radial distance of closest approach (DCA$_R$) by projecting the particle 
track from the FVTX onto a plane in the z-axis at the primary vertex, as 
identified by the VTX. Precise measurement of DCA$_R$ enables a 
statistical separation of muons from short-lived particles, such as 
heavy-flavor mesons, and long-lived particles like pions and kaons, 
thereby significantly improving the signal-to-background ratio in the 
low mass region. The FVTX also provides precise tracking for charged 
particles entering the muon spectrometer before they undergo multiple 
scattering in the hadron absorber. Located downstream of the FVTX with 
respect to the interaction region, the hadron absorber comprises 
layers of copper, iron, and stainless steel, with a total thickness of 
7.2 interaction lengths. It efficiently attenuates hadrons before they 
reach the muon arm, significantly reducing the hadronic background for 
muon-related measurements.

The absorber is followed by the muon spectrometer, which is composed of 
a muon tracker (MuTr) and a muon identifier (MuID). The MuTr, which is 
made up of three stations of cathode strip chambers placed within a 
magnet with a radial field integral of $0.72$ T$\cdot$m. The MuID 
comprises five alternating layers of steel absorbers and Iarocci 
tubes. The overall momentum resolution, $\delta p/p$, for particles 
within the analyzed momentum range is $\approx$5\%, independent of 
momentum, and is mainly constrained by multiple scattering.

In addition, two beam-beam counters (BBCs) are located on both sides of 
the interaction point and cover the pseudorapidity range 
$3.1<|\eta|<3.9$. These BBC detectors comprise 128 identical quartz 
photomultiplier tubes and function as a \v{C}erenkov array. The BBC is 
used to measure the beam luminosity and form a minimum-bias (MB) 
trigger. It is also used to determine the vertex position along the 
direction of the beam and classify event centrality based on the total 
charge recorded in the BBC. To categorize centrality, a Monte Carlo 
Glauber model calculation is employed. The model takes inputs such as 
the inelastic nucleon-nucleon cross section and the nuclear charge 
density to simulate the probability of collision between nuclei based on 
the nucleon-nucleon inelastic cross section. The total charge detected 
by the BBC is used to define the centrality classes, and Glauber model 
provides the mean number of binary (nucleon-nucleon) collisions, \Ncoll, 
for each centrality range~\cite{PhysRevC.84.054912,PhysRevC.90.034902}.

%%%%%%%%%%%%%%%%%%%%%%%%%%%%%%%%%%%%%%%%%%%%%%%%%%%%%%%%%%%%%%%%%%%%%

\section{DATA ANALYSIS}
\subsection{Dataset and quality cuts}
\label{subsect:cuts}

This section provides the details of the LVM measurements in the dimuon 
decay channel. The \auau data set used here was collected in 2014 from 
\auau collisions at \sqsntwo, using a MB trigger requiring at least two 
hits in each of the BBC detectors. The MB trigger captures $92\pm2$\% of 
the total \auau inelastic cross section. The \pp data set for this 
analysis was recorded in 2015 from \pp collisions at \sqs = 200 GeV, 
using a MB trigger that required at least one hit in each of the BBC 
detectors. In addition, the MuID Level-1 dimuon trigger was applied, 
which required at least two tracks to penetrate through the MuID to its 
last layer.

Muons are identified by requiring that a detected particle penetrates 
several layers of absorber material in the MuID, which drastically 
reduces the hadron contribution although some hadrons may punch through. 
The particles that pass this threshold are then matched with the tracks 
reconstructed by the MuTr. The FVTX detector provides additional space 
points near the collision vertex before the particle begins its 
trajectory through the muon-arm absorbers, improving the mass 
resolution.

A set of selections is applied to the data to isolate high-quality muon 
candidates and enhance the signal-to-background ratio. Event selection 
requires the BBC collision z-vertex to be reconstructed within $\pm10$ 
cm from the center of the interaction region along the beam axis. MuTr 
tracks are matched to MuID tracks at the first MuID layer in both 
position and angle, and only dimuon candidates in which both tracks 
penetrate the final MuID layer are considered. Furthermore, the track 
must have a minimum number of possible hits in both MuTr and MuID, and a 
maximum allowed $\chi^2$ is applied to both the track and vertex 
determination. A minimum single-muon momentum along the beam axis, 
$p_z$, is required, which is reconstructed and energy loss corrected at 
the collision vertex. Furthermore, MuTr + MuID tracks are matched with 
FVTX tracks and a strict cut on DCA$_R$ is applied. Finally, this 
analysis is restricted to the dimuon rapidity region of $1.2<|y|<2.2$ 
and \pt range of 2.5--5.0~GeV/$c$. The \pt limitation is due to the large 
background and limited acceptance at low \pt and low statistics at high 
\pt.

\subsection{Detector acceptance and reconstruction efficiency}
\label{subsect:acc_eff}

The acceptance and reconstruction efficiency (\accEff) is determined 
using Monte Carlo (MC) simulation. The \accEff is defined by the number 
of dimuons reconstructed in the muon spectrometers relative to the 
number of dimuons generated in the same kinematic region. The kinematic 
distributions of {\sc pythia}\footnote{We used {\sc pythia6} (ver 
6.421), with parton distribution functions given by CTEQ6LL. The 
following parameters were modified: MSEL = 0, MSUB(86) = 1, PARP(91) = 
2.1, MSTP(51) = 10041, MDME(858,1) = 0, MDME(859,1) = 1, MDME(860,1) = 
0, and Tune A.}~\cite{SJOSTRAND2001238} generated \pt, rapidity, and LVM 
mass shape were used as input into a full PHENIX {\sc geant4} 
simulation~\cite{AGOSTINELLI2003250}. The \pt, rapidity and vertex 
distributions were tuned such that the reconstructed distributions match 
those of the data. Variations within the uncertainties of the data are 
taken as systematic uncertainty.

The detector response in the simulation is tuned to a set of 
characteristics (dead and hot channel maps, gains, noise, etc.) that 
describes the performance of each detector subsystem. The simulated 
events are further embedded with real data to account for the effects of 
detector occupancy, noise and fake tracks, and then reconstructed in the 
same manner as the real data. As a cross-check, the calculated \Jpsi 
invariant yield, after applying the \accEff correction, showed good 
agreement with the published result~\cite{PhysRevLett.98.232301} within 
the statistical uncertainties across all \pt and centrality bins.

%%%%%%%%%%%%%%%%%%%%%%%%%%%%%%%%%%%%%%%%%%%%%%%%%%%%%%%%%%%%%% Fig_2
\begin{figure}
\centering
\includegraphics[width=1.0\linewidth]{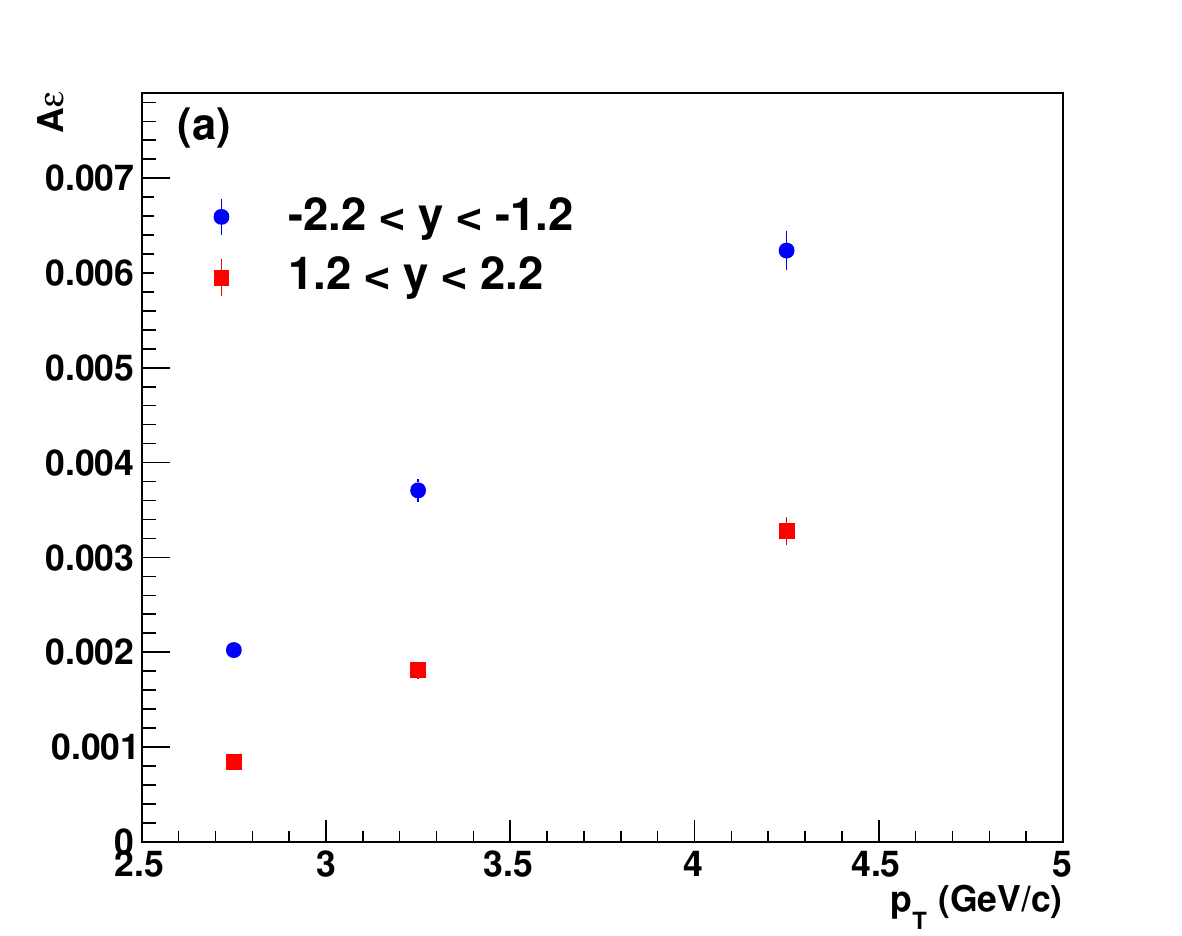}
\includegraphics[width=1.0\linewidth]{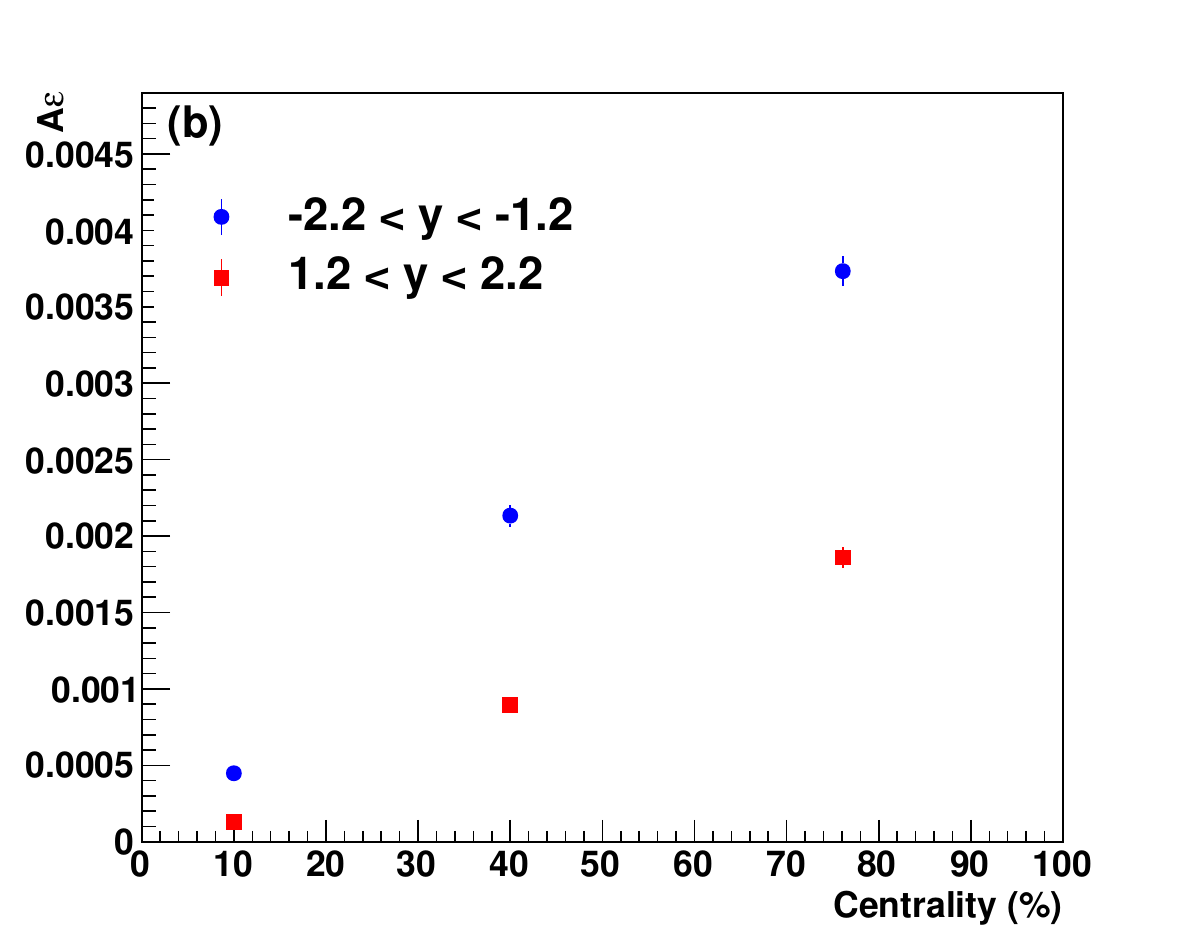}
\caption{\label{fig:AccEff} \accEff of dimuon pairs in \auau collisions 
as a function of (a) \pt, (b) centrality.}
\end{figure}

Figure~\ref{fig:AccEff} shows the \accEff of dimuon pairs as a function 
of (a) \pt, (b) centrality (\%) for the forward (north) and backward 
(south) rapidities, covering the rapidity ranges $1.2<y<2.2$ and 
$-2.2<y<-1.2$, respectively. The relative difference in \accEff between 
the two spectrometers is due to the different detection efficiencies of 
the MuTr and MuID systems and different amounts of absorber material.

\subsection{Raw yield extraction}
\label{subsec:yieldext}

The invariant mass distribution is constructed by combining muon 
candidate tracks of opposite charge. In the mass range 
$0.3<m_{\mu\mu}<2.5$~GeV/$c^2$, this unlike-sign invariant mass spectrum 
includes $\phi$, $\rho$, and $\omega$ mesons, as well as uncorrelated 
and correlated backgrounds. The uncorrelated backgrounds arise from 
random combinatorial associations of muon candidates, while the 
correlated backgrounds result from the following processes: open charm 
decay (e.g., $D\bar{D}$ pairs where both decay semileptonically to 
muons), open beauty decay, $\eta$ meson and $\omega$ meson Dalitz 
decays, or the Drell-Yan process.

The uncorrelated combinatorial background is addressed using two 
distinct methods: (1) like-sign dimuons and (2) event mixing. The 
like-sign dimuon background is constructed by pairing muon candidate 
tracks of the same charge, and is used to model the uncorrelated 
background under the assumption that like-sign dimuon pairs arise purely 
from combinatorial processes, with no correlation between single muons. 
In the event-mixing approach, muons from different events are randomly 
paired to generate a background distribution of uncorrelated dimuon 
pairs. Events are mixed with four previous events within the same 
2\%~centrality and 1-cm~z-vertex bins to minimize systematic 
uncertainties.  Detailed descriptions of both techniques can be found in 
Ref.~\cite{PhysRevC.93.024904}. Although the uncorrelated background 
distributions were consistent between the two methods, the event-mixing 
technique is employed in this analysis due to the statistical 
limitations of the like-sign method.

To extract the LVM signal, the unlike-sign invariant mass spectrum is 
fitted with a combination of two Gaussian functions, a Breit-Wigner 
function convolved with a Gaussian, and fitting functions for both 
correlated and uncorrelated backgrounds. The two Gaussian functions are 
used to fit the $\omega$ and $\phi$ mesons, while the Breit-Wigner 
convolved with a Gaussian is used to fit the $\rho$ meson. The fit 
parameters are fixed to the world average values for the masses of the 
three vector mesons~\cite{PhysRevD.86.010001} while the widths of the 
Gaussian distributions account for the detector mass resolution and are 
constrained by the values obtained from simulation. Because the 
invariant mass peaks of $\omega$ and $\rho$ mesons cannot be resolved, 
the combined yield of these two mesons is extracted. The ratio of the 
$\omega$ and $\rho$ mesons, $N_\rho/N_\omega$, is set to 0.58, derived 
as the ratio of their corresponding production cross sections, 
$\sigma_\rho/\sigma_\omega = 1.15 \pm 0.15$, consistent with values 
found in jet fragmentation~\cite{J.Phys.G33.1}, multiplied by the ratio 
of their branching ratios~\cite{PhysRevD.86.010001,PhysRevD.90.052002}.

To describe the uncorrelated background in the unlike-sign invariant 
mass spectrum, the mixed-event distribution is fitted with a modified 
Hagedorn 
function~\cite{PhysRevC.84.044905,PhysRevD.99.072003,PhysRevC.102.014902}:

\begin{equation}\label{eqn:corrbkgd}
	 \frac{d^2N}{dm_{\mu\mu}dp_T} = \frac{p_0}
     {[\exp(-p_1m_{\mu\mu}-p_2m^2_{\mu\mu})+m_{\mu\mu}/p_3]^{p_4}},
 \end{equation}
where $m_{\mu\mu}$ is the reconstructed dimuon mass, $p_0$ is a 
normalization parameter, $p_4$ is the high mass tail parameter, and 
$p_1$, $p_2$, and $p_3$ are additional fit parameters. To assess the 
stability of the modified Hagedorn fit, the mixed-event distribution was 
fitted by a polynomial of $4^{th}$ order and a modified Hagedorn 
function is used to fit the unlike-sign dimuon spectrum in the 
nonresonance region. The results of all three fits were consistent 
within uncertainties and the variations are taken as a systematic 
uncertainty on the yield extraction. The correlated background is well 
described by the function in Eq.~\ref{eqn:corrbkgd},

 \begin{equation}\label{eqn:corrbkgd2}
	 f(m_{\mu\mu}) = a\exp(bm_{\mu\mu}),
 \end{equation}
where $a$ and $b$ are free parameters of the fit $f(m_{\mu\mu})$.

%%%%%%%%%%%%%%%%%%%%%%%%%%%%%%%%%%%%%%%%%%%%%%%%%%%%%%%%%%%%%% Fig_3
\begin{figure*}[hbt]
    \includegraphics[width=0.99\linewidth]{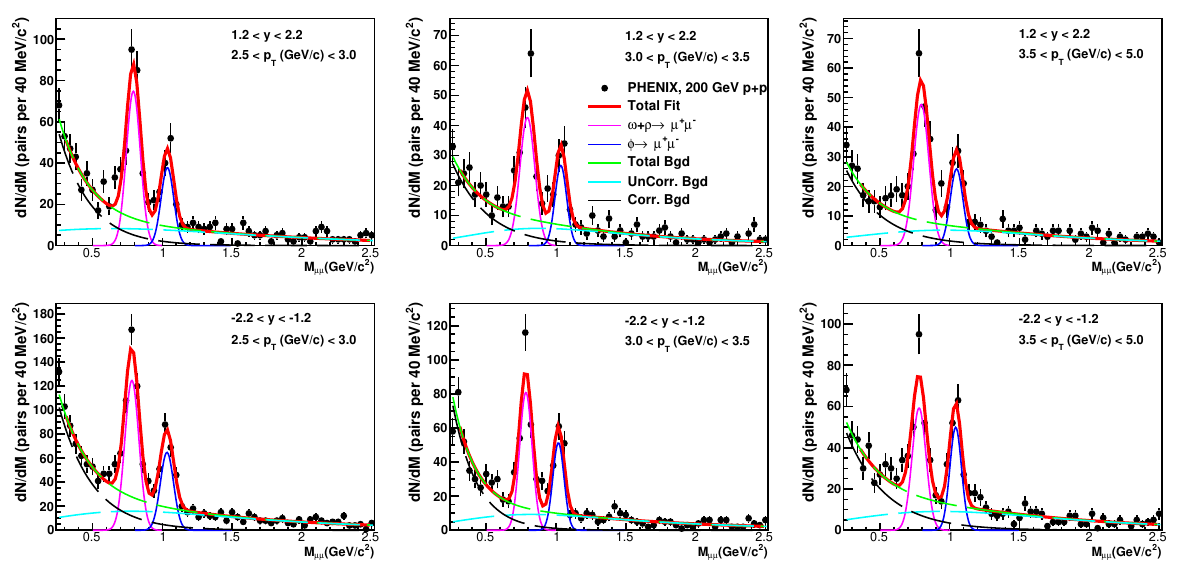}
    \caption{The fitted raw unlike-sign dimuon spectra (black points) in 
\pp collisions are shown for the different \pt bins at forward (top) and 
backward (bottom) rapidities. The panels also show the fits for the 
uncorrelated background (dashed cyan), correlated background (dashed 
black), \omgRho mesons (solid magenta), $\phi$ meson (solid blue), and 
the total fit (solid red).}
\label{fig:ppRawYdFits}
\end{figure*}

Each rapidity region $(1.2<|y|<2.2)$ in the \pp data set is sorted into the 
following \pt bins: $[2.5, 3.0]$, $[3.0, 3.5]$, and $[3.5, 6.0]$ GeV/$c$. 
Figure~\ref{fig:ppRawYdFits} shows the results of fitting the dimuon spectra 
for each of these \pt bins. The fit function in Fig.~\ref{fig:ppRawYdFits} does 
not adequately capture the \omgRho signal peak. As a result, the unlike-sign 
spectra were integrated over the \omgRho and $\phi$ signal regions after 
subtracting the total background fit function to obtain the raw yields. The 
integral and fit values were consistent for the $\phi$ meson, while for \omgRho 
mesons, the fit values were up to 8\% lower than the integral values in some 
cases. Therefore, the integral values were adopted.

In the \auau data set, the statistics are insufficient to sort the data 
into \pt or centrality bins for each arm individually. However, due to 
the symmetry of the \auau collisions, the invariant yields for each arm 
are found to be consistent, with the means and widths of the fits for 
both $\phi$ and \omgRho mesons being consistent between arms. Therefore, 
the raw yield as a function of \pt and as a function of centrality can 
be extracted for the combined sum of the north and south arms. Because the 
north and south arms have different \accEff values, as shown in 
Fig.~\ref{fig:AccEff}, a simple sum cannot be used. Instead, the north 
arm is scaled by $A\varepsilon_{south}/A\varepsilon_{north}$, as a 
function of \pt and centrality, before being added to the south arm. The 
resulting yields are then weighted by $A\varepsilon_{south}$ when 
calculating the invariant yield.

%%%%%%%%%%%%%%%%%%%%%%%%%%%%%%%%%%%%%%%%%%%%%%%%%%%%%%%%%%%%%% Fig_3
\begin{figure*}[hbt]
    \includegraphics[width=0.99\linewidth]{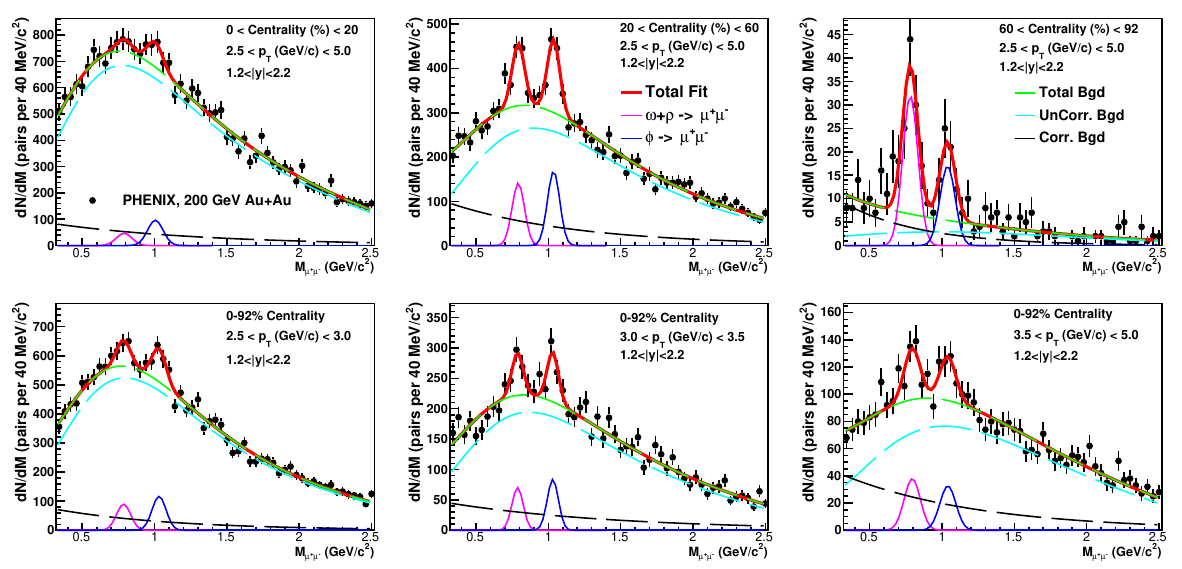}
    \caption{The fitted raw unlike-sign dimuon spectra (black points) in 
\auau collisions are shown for different centrality classes (top) and 
\pt bins (bottom). The panels also show the fits for the uncorrelated 
background (dashed cyan), correlated background (dashed black), \omgRho 
mesons (solid magenta), $\phi$ meson (solid blue), and the total fit 
(solid red).}
\label{fig:rawYdFits}
\end{figure*}

Similar to the \pp data set, the \pt distribution in the \auau data set 
is sorted into the following bins: $[2.5, 3.0], [3.0, 3.5]$, and $[3.5, 
6.0]$~GeV/$c$, while the centrality distribution is sorted into the bins: 
$[0, 20], [20, 60],$ and $[60, 92]$. Figure~\ref{fig:rawYdFits} shows the 
results of fitting the dimuon spectra for different centrality and \pt 
bins in case of the \auau data set. The fits in this figure include a 
modified Hagedorn to describe the uncorrelated background. As discussed 
earlier, two additional descriptions are used in extracting the LVM 
yields. The LVM yields were consistent among all three fits within the 
uncertainties. Therefore, the best estimate for the invariant yield is 
taken as the RMS of all the fit methods, with its uncertainty calculated 
as the RMS of their individual uncertainties. The variations are 
considered as the systematic uncertainty for the yield extraction.

\subsection{Calculation of invariant yields and nuclear-modification factors}

The dimoun LVM invariant yield in a given \pt and centrality bin is
 \begin{equation}\label{RAA}
\frac{B_{\mu\mu}}{2\pi p_T}\frac{d^2N}{dydp_T}=
\frac{1}{2\pi p_T}\frac{1}{\Delta y\Delta p_T}\frac{N_{\omega+\rho (\phi)}}{A\varepsilon N_{\rm evt}},
 \end{equation}
where $B_{\mu\mu}$ is the branching ratio to dimuons, $N_{\omega+\rho 
(\phi)}$ is the number of observed mesons, $N_{\rm evt}$ is the number of MB 
events sampled in the given centrality bin, $\Delta y$ is the width of 
the rapidity bin, and $\Delta p_T$ is the width of the \pt bin. \accEff 
is the acceptance and reconstruction efficiency.\\ In \pp collisions, 
the differential cross section, $d^2\sigma/dydp_T$, is evaluated 
according to the following relation:

 \begin{equation}\label{diffxs}
\frac{d^2\sigma}{dydp_T} = \frac{d^2N}{dydp_T}\sigma_{\rm tot},
 \end{equation}
where $\sigma_{\rm tot}=\sigma_{\rm BBC}/\varepsilon^{\rm BBC}_{\rm MB}$ is the total 
cross section. $\sigma_{\rm BBC}$ is the BBC cross section, $23.0 \pm 2.2$ 
mb at \sqs = 200 GeV, which is determined from the van der Meer scan 
technique, and $\varepsilon^{\rm BBC}_{\rm MB}$ is MB trigger efficiency, 
0.545~\cite{PhysRevD.79.012003}.

To gain insight into nuclear medium effects and particle production 
mechanisms in \auau collisions, the ratio of the LVM yields in \auau 
collisions to \pp collisions scaled by \Ncoll for that centrality bin in 
the \auau system~\cite{PhysRevLett.91.072301}, is calculated as

 \begin{equation}\label{RxA}
	 R_{AA}=\frac{d^2N_{AA}/dydp_T}{\langle N_{\rm coll}\rangle\times d^2N_{pp}/dydp_T},
 \end{equation}
where $R_{AA}$ is the nuclear-modification factor, $d^2N_{AA}/dydp_T$ is 
the per-event yield of particle production in \auau collisions and 
$d^2N_{pp}/dydp_T$ is the per event yield of the same process in \pp 
collisions~\cite{PhysRevD.90.052002}.

\subsection{Systematic uncertainties}

The systematic uncertainties are categorized into three types according 
to their impact on the measured results. All uncertainties are expressed 
as one standard deviation.

Type-A: point-to-point uncorrelated uncertainties, meaning that the data 
points can vary independently of each other. These uncertainties are 
combined in quadrature with statistical uncertainties. To account for 
the variation in yield when different functions are used to fit the 
uncorrelated background, a 6\% uncertainty is assigned in signal 
extraction for the \auau data set and for the \pp data set a 5.0\% 
(4.5\%) uncertainty is assigned for the north (south) arm.

Type-B: point-to-point correlated uncertainties which allow the data 
points to move coherently within the quoted range to some degree. These 
systematic uncertainties include a 4\% uncertainty from MuID tube 
efficiency and a 2\% from MuTr overall efficiency. Furthermore, the 
systematic uncertainty in \accEff due to the input \pt is 6.3\%. In the 
case of the \auau data set, the systematic uncertainties related to the 
azimuthal distribution of the muon tracks are 10.4\% for the north arm 
and 6.7\% for the south arm. Because invariant yields and nuclear 
modification factors are calculated by adding the north and south arms, 
the azimuthal distribution uncertainties are added in quadrature, 
resulting in a total uncertainty of 12.4\%. In the case of the \pp data, 
the systematic uncertainty related to the azimuthal distribution of the 
muon tracks is 7.5\% (6.4\%) for the north (south) arm. The Type-B 
systematic uncertainties are added in quadrature. For the \auau dataset, 
these uncertainties total 14.6\%. In the case of the \pp dataset, they 
amount to 10.3\% for the north arm and 9.6\% for the south arm. These 
uncertainties are shown as boxes around the data points.

Type-C: an overall normalization uncertainty of 10\% for \pp is assigned 
to the BBC bias correction uncertainties~\cite{PhysRevC.87.044909}. No 
correction is applied for \auau collisions. All systematic uncertainties 
included in the invariant calculations are summarized in 
Table~\ref{tab:sysUncer}.

%%%%%%%%%%%%%%%%%%%%%%%%%%%%%%%%%%%%%%%Table 1
\begin{table}[ht!]
\caption{\label{tab:sysUncer} 
Systematic uncertainties included in the invariant yield calculations.}
\begin{ruledtabular}
\begin{tabular}{cccc}
Type & Origin & \auau & \pp\\
     &        &       & north (south) \\
\hline
A & Signal extraction & 6\% & 5\% (4.5\%)\\
B & MuID hit efficiency & 4\% & 4\% \\
B & MuTr hit efficiency & 2\% & 2\% \\
B & \accEff \pt & 6.3\% & 5.5\% \\
B & \accEff $\varphi$ distribution & 12.4\% & 7.5\% (6.4\%) \\
B & Quadratic sum & 14.6\% & 10.3\% (9.6\%)\\
C & MB trigger efficiency & -- & 10\%\\
\end{tabular}
\end{ruledtabular}
\end{table}

%%%%%%%%%%%%%%%%%%%%%%%%%%%%%%%%%%%%%%%%%%%%%%%%%%%%%%%%%%%%%% Fig_5
\begin{figure*}[hbt]
\begin{minipage}{0.48\linewidth}
    \includegraphics[width=1.0\linewidth]{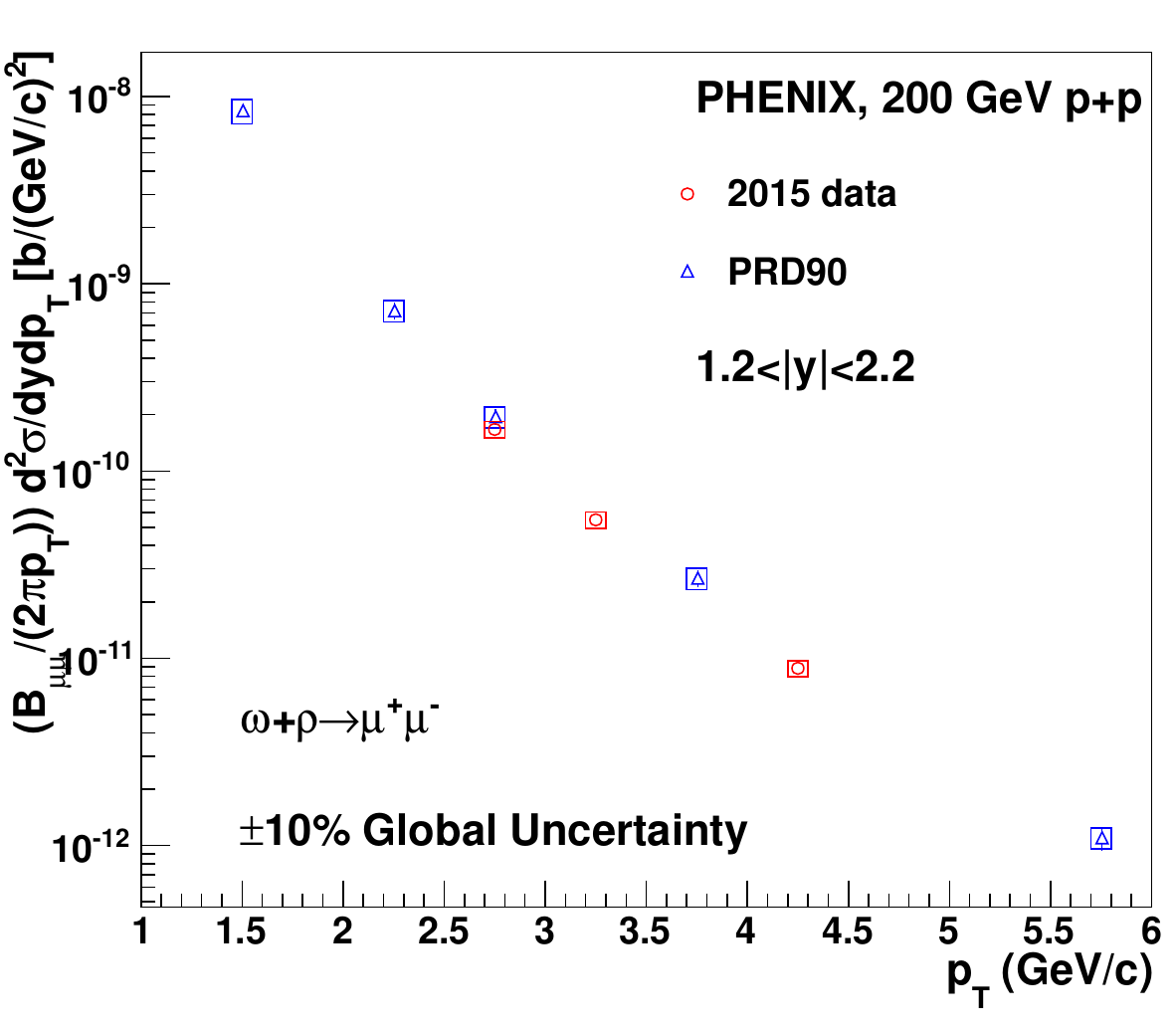}
    \caption{The differential cross sections of \omgRho mesons as a 
function of \pt in \pp collisions for the data sets collected in 2015 
(empty red circles) and 2009 (empty blue 
triangles)~\cite{PhysRevD.90.052002}. The systematic uncertainties of 
type-A are combined with statistical uncertainties in quadrature and are 
represented as error bars. Type-B systematic uncertainties are displayed 
as boxes around the data points.}
\label{fig:omgXS}
\end{minipage}
%\end{figure}
%%%%%%%%%%%%%%%%%%%%%%%%%%%%%%%%%%%%%%%%%%%%%%%%%%%%%%%%%%%%%% Fig_6
%\begin{figure}[hbt]
\hspace{0.3cm}
\begin{minipage}{0.48\linewidth}
    \includegraphics[width=1.0\linewidth]{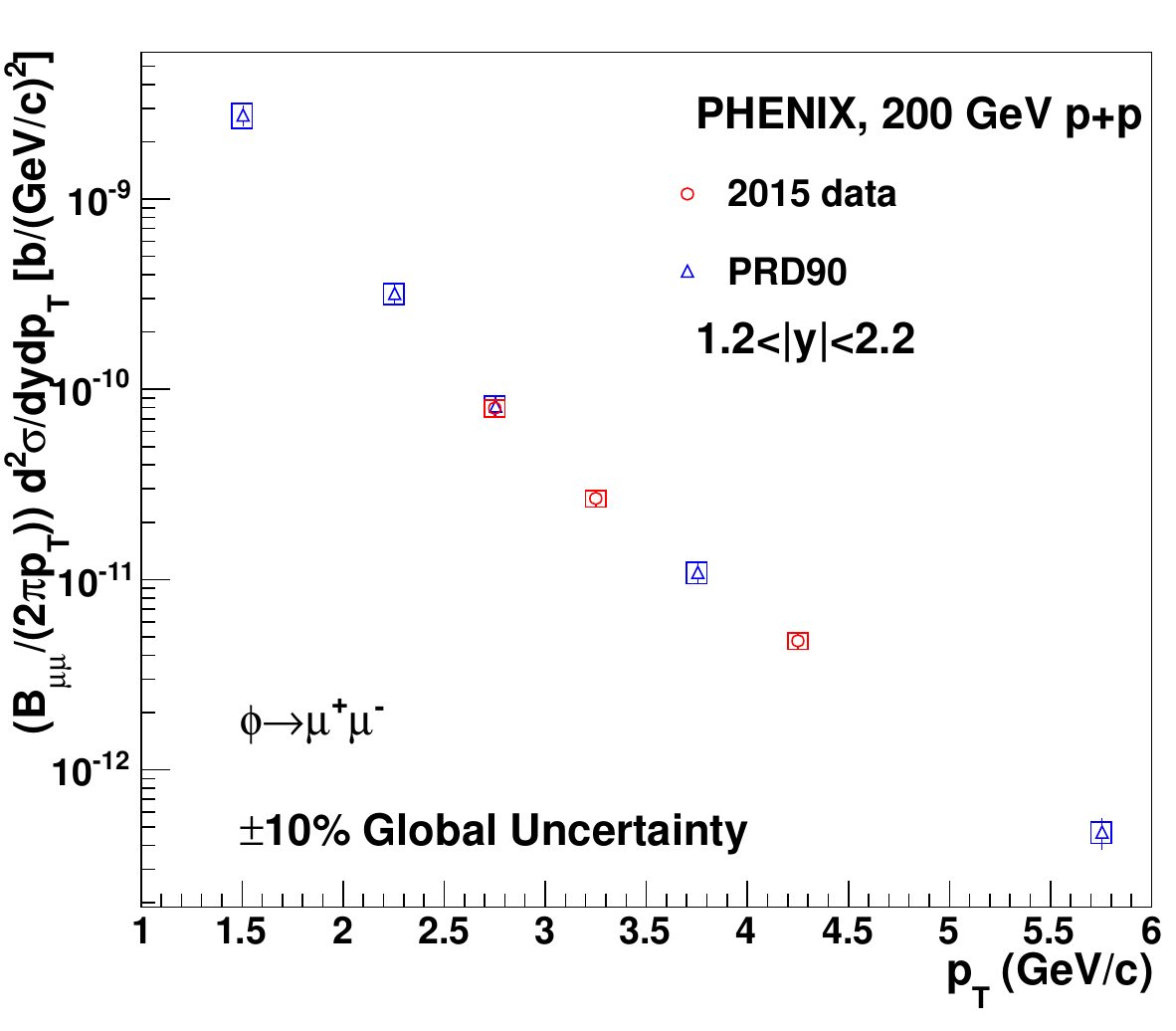}
    \caption{The differential cross sections of $\phi$ meson as a 
function of \pt in \pp collisions for the data sets collected in 2015 
(empty red circles) and 2009 (empty blue 
triangles)~\cite{PhysRevD.90.052002}. The systematic uncertainties of 
type-A are combined with statistical uncertainties in quadrature and are 
represented as error bars. Type-B systematic uncertainties are displayed 
as boxes around the data points.}
\label{fig:phiXS}
\end{minipage}
\end{figure*}

For the nuclear-modification factor, the type-A systematic uncertainties 
are included in the statistical uncertainties of the \auau and \pp 
invariant yields, as mentioned earlier. The systematic uncertainties 
that include MuID and MuTr efficiencies are the same between the 
invariant yields of \auau and \pp and cancel out. A 9.3\% (8.4\%) 
systematic uncertainty in \accEff for the north (south) arm, which is 
carried over from \pp, is added in quadrature to the type-B systematic 
uncertainties listed in Table~\ref{tab:sysUncer}. Type-C systematic 
uncertainties for \raa are calculated as the quadratic sum of the Type-C 
systematic uncertainties from the \pp collisions and the uncertainty in 
\Ncoll. The latter contributes an uncertainty of 9\%, based on the 
Glauber model calculations for \auau collisions averaged over all 
centralities. However, when \raa is presented as a function of \Npart, 
the uncertainty associated with \Ncoll is instead included as a Type-B 
uncertainty, summed in quadrature with other Type-B contributions.

%%%%%%%%%%%%%%%%%%%%%%%%%%%%%%%%%%%%%%%%%%%%%%%%%%%%%%%%%%%%%%%%%%%%%%%%%%%
\section{RESULTS}

The differential cross sections of \omgRho and $\phi$ mesons as a 
function of \pt in \pp collisions at \sqs = 200 GeV are presented in 
Figs.~\ref{fig:omgXS} and~\ref{fig:phiXS}, respectively. Both figures 
show comparisons between previously published PHENIX 
data~\cite{PhysRevD.90.052002} and current measurements from the data 
collected in 2015, and they are consistent within uncertainties. The 
invariant yields of \omgRho and $\phi$ mesons in \pp collisions at 
\sqs=200 GeV, obtained from the data collected in 2015, serve as a 
reference for the measurements of the nuclear-modification factor.

%\clearpage

%%%%%%%%%%%%%%%%%%%%%%%%%%%%%%%%%%%%%%%%%%%%%%%%%%%%%%%%%%%%%% Fig_7
\begin{figure*}[hbt]
\begin{minipage}{0.48\linewidth}
    \includegraphics[width=1.0\linewidth]{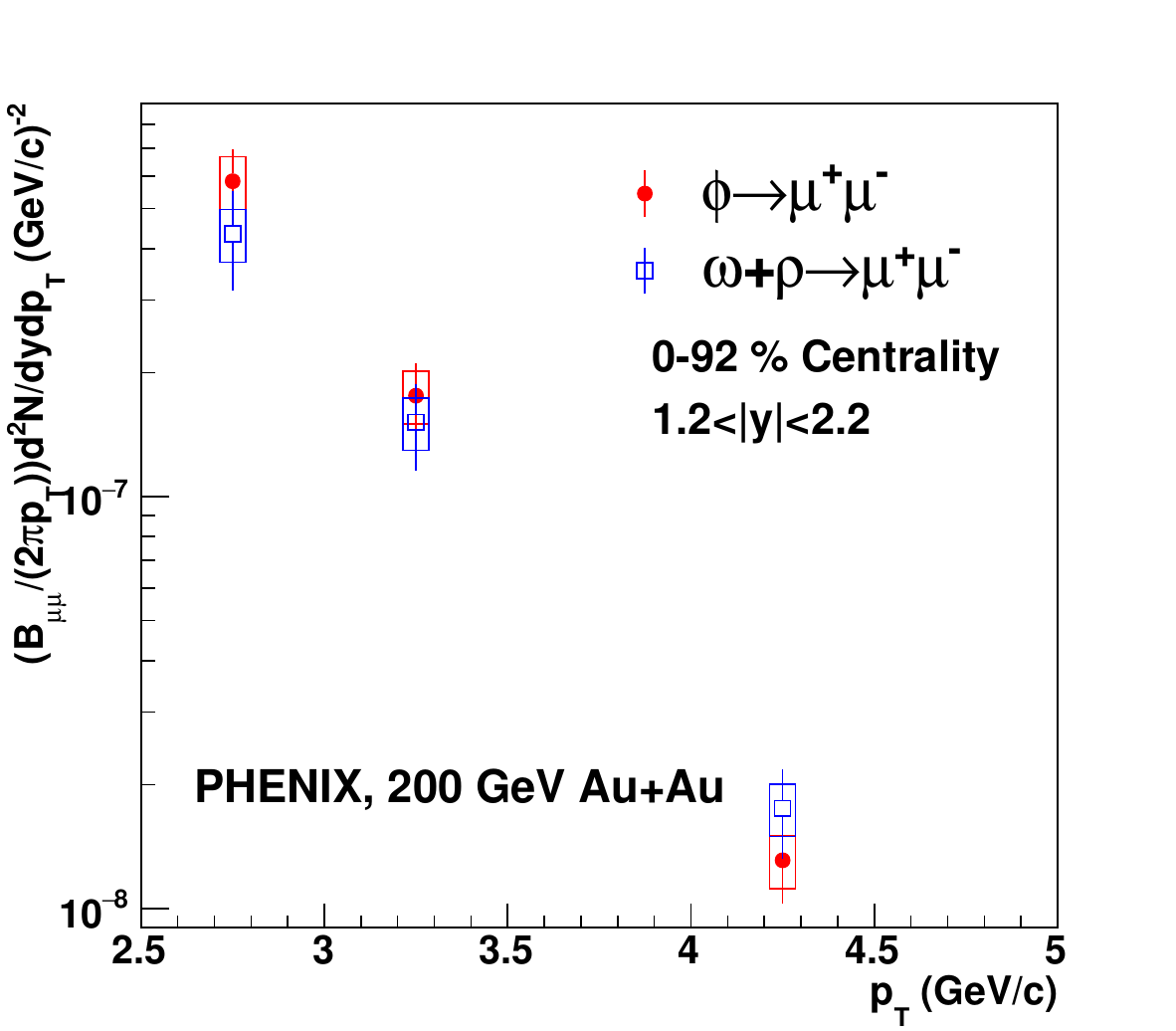}
    \caption{The invariant yields of $\phi$ (solid red circles) and 
\omgRho (empty blue squares) mesons as a function of \pt in Au$+$Au 
collisions. The systematic uncertainties of type-A are combined with 
statistical uncertainties in quadrature and are represented as error 
bars. Type-B systematic uncertainties are displayed as boxes around the 
data points.}
\label{fig:invYdpt}
\end{minipage}
%\end{figure}
%%%%%%%%%%%%%%%%%%%%%%%%%%%%%%%%%%%%%%%%%%%%%%%%%%%%%%%%%%%%%% Fig_8
%\begin{figure}[hbt]
\hspace{0.3cm}
\begin{minipage}{0.48\linewidth}
    \includegraphics[width=1.0\linewidth]{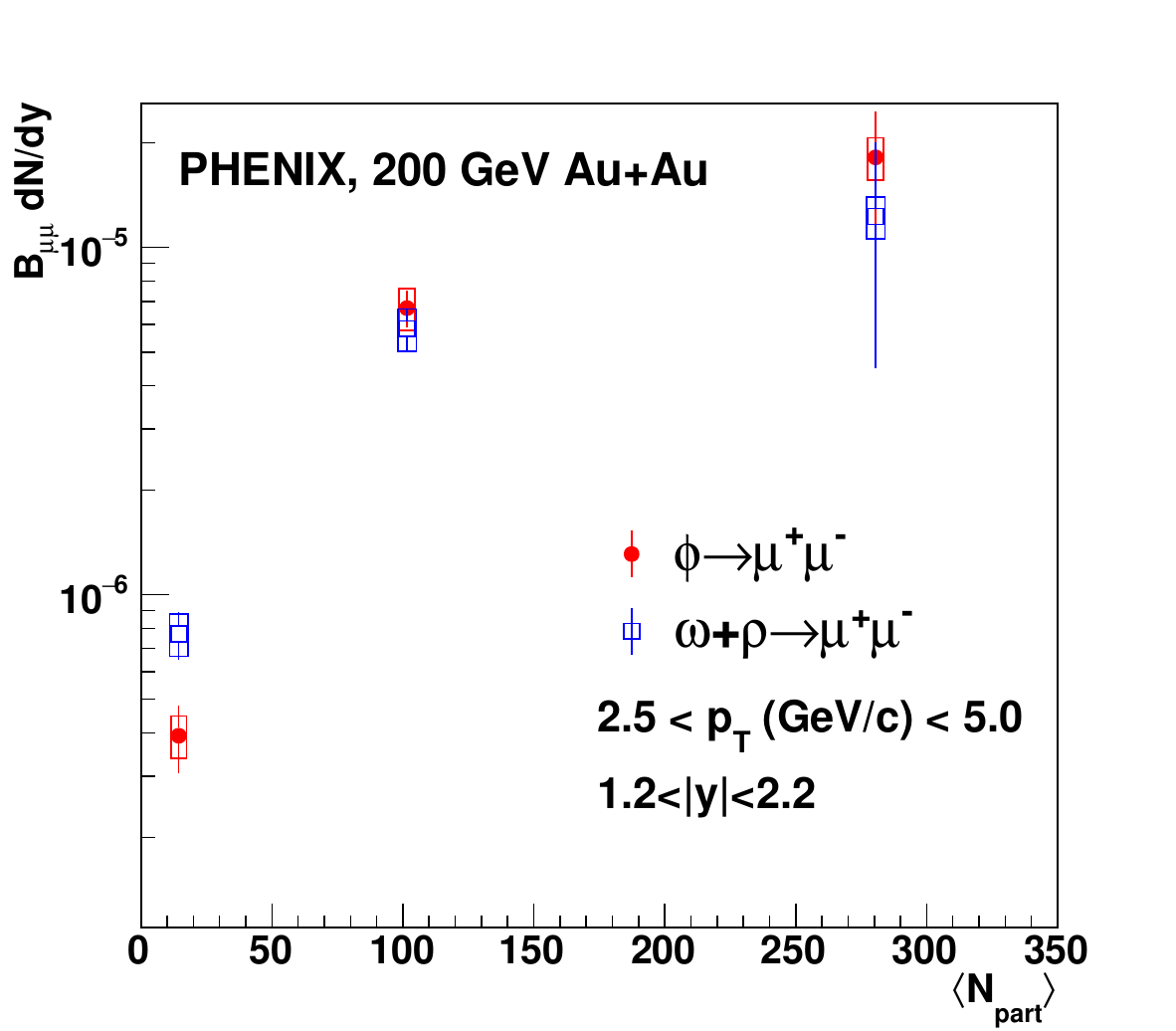}
    \caption{The invariant yields of $\phi$ (solid red circles) and 
\omgRho (empty blue squares) mesons as a function of \Npart in Au$+$Au 
collisions. The systematic uncertainties of type-A are combined with 
statistical uncertainties in quadrature and are represented as error 
bars. Type-B systematic uncertainties are displayed as boxes around the 
data points.}
\label{fig:invYdnpart}%
\end{minipage}
\end{figure*}

%%%%%%%%%%%%%%%%%%%%%%%%%%%%%%%%%%%%%%%%%%%%%%%%%%%%%%%%%%%%%% Fig_9
\begin{figure*}[htbp]
\begin{minipage}{0.48\linewidth}
    \includegraphics[width=1.0\linewidth]{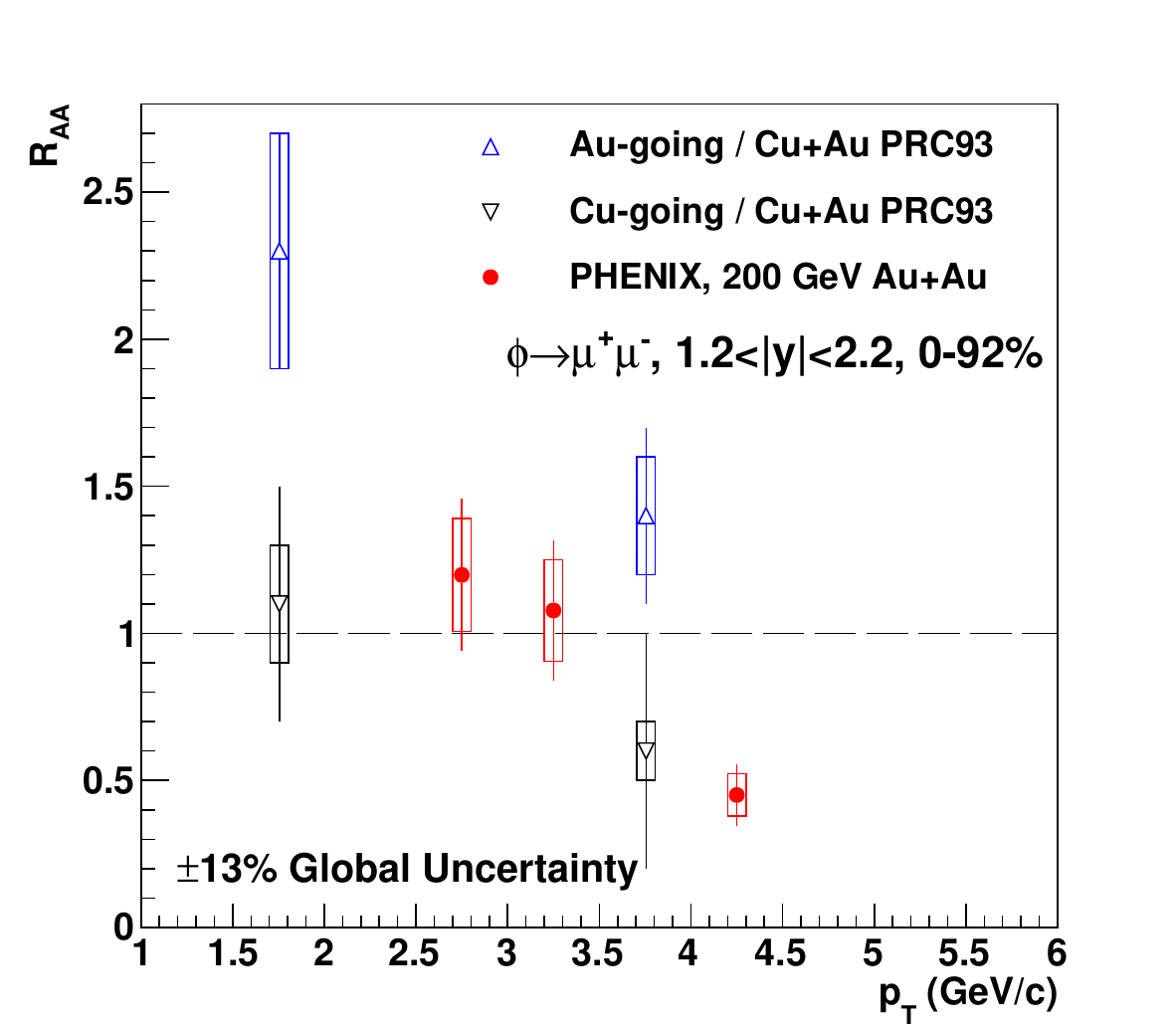}
    \caption{The \raa of $\phi$ meson as a function of \pt for $1.2<|y|<2.2$ in 
\auau collisions at \sqsntwo, compared with \cuau 
collisions~\protect\cite{PhysRevC.93.024904}. The systematic uncertainties of 
type-A are combined with statistical uncertainties in quadrature and are
 displayed as error bars. Type-B systematic uncertainties are shown 
as boxes around the data points.}
\label{fig:phiRaa_pt}
\end{minipage}
%\end{figure}
%%%%%%%%%%%%%%%%%%%%%%%%%%%%%%%%%%%%%%%%%%%%%%%%%%%%%%%%%%%%%% Fig_10
%\begin{figure}[htbp]
\hspace{0.3cm}
\begin{minipage}{0.48\linewidth}
    \includegraphics[width=1.0\linewidth]{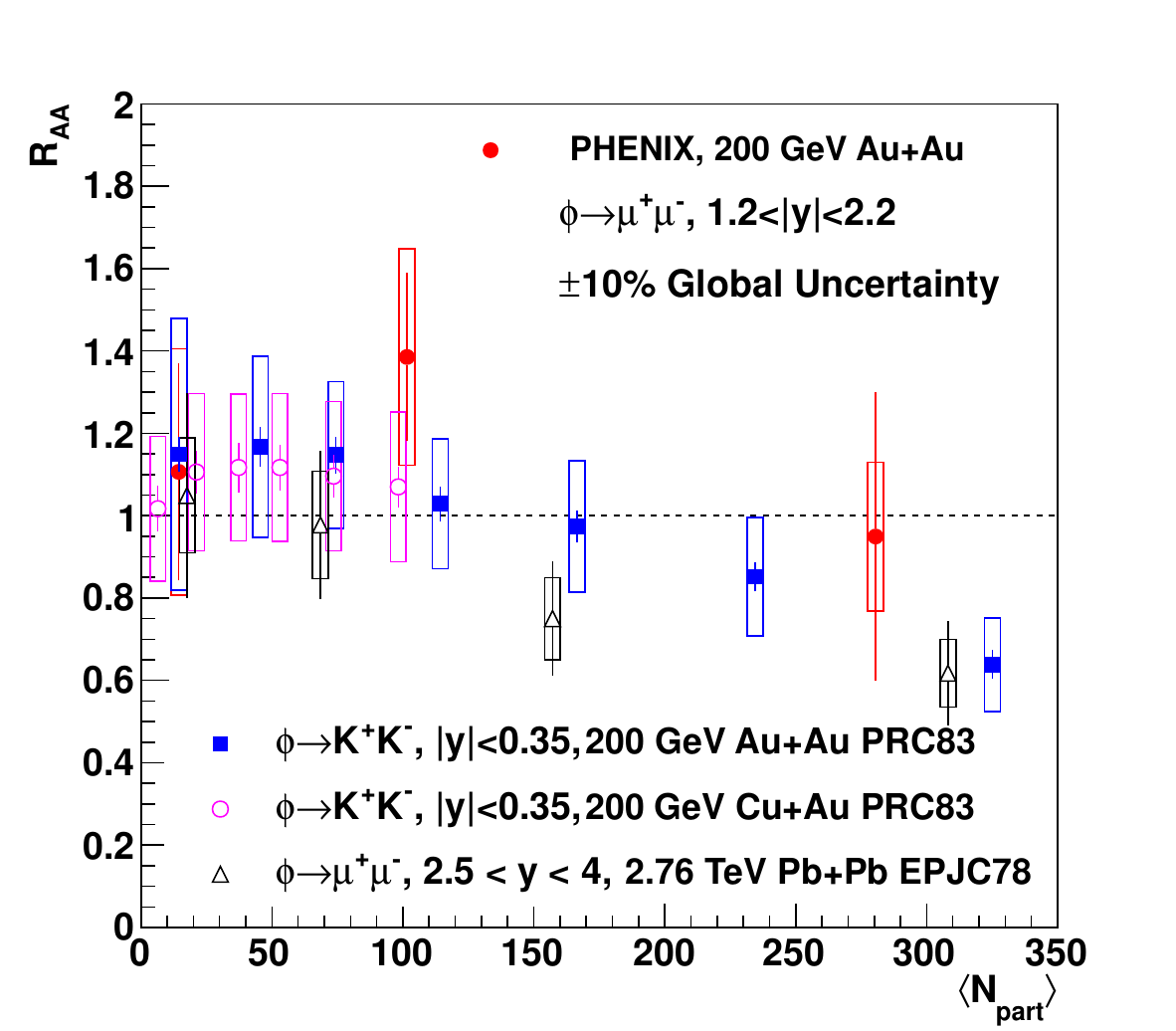}
    \caption{The \raa of $\phi$ meson as a function of \Npart for 
$1.2<|y|<2.2$ in \auau collisions at \sqsntwo, compared with the PHENIX 
measurements at $|y|<0.35$ in \auau and \cuau collisions at 
\sqsntwo~\protect\cite{PhysRevC.83.024909}. Results at the LHC from ALICE at 
$2.5<y<4$ in Pb$+$Pb collisions at 2.76 TeV are also shown for
comparison~\protect\cite{Acharya2018}.}
\label{fig:phiRaa_npt}
\end{minipage}
\end{figure*}

Figures~\ref{fig:invYdpt} and~\ref{fig:invYdnpart} show the LVM 
invariant yields in \auau as a function of \pt, integrated over 
0\%--92\% centrality, and as a function of the average number of 
participating nucleons (\Npart), respectively. The $\phi$ meson 
distribution is softer compared to the \omgRho mesons distribution 
within the studied \pt range, and its yield is lower than the \omgRho 
yield in the most peripheral collisions.

Figure~\ref{fig:phiRaa_pt} shows the \raa for the $\phi$ meson in \auau 
collisions integrated over all centralities as a function of \pt, 
measured over the rapidity range $1.2 < |y| < 2.2$. The \raa shows a 
hint of enhancement at intermediate \pt, though it remains consistent 
with unity within uncertainties, and exhibits significant suppression at 
higher \pt. Additionally, Fig.~\ref{fig:phiRaa_pt} shows the \raa for 
the $\phi$ meson in \cuau collisions, measured across all centralities 
and at the same rapidity~\cite{PhysRevC.93.024904}. The Au-going result 
is consistent with this measurement within uncertainties. Similar 
nuclear modification pattern was observed at 
midrapidity~\cite{PhysRevC.107.014907}.

Figure~\ref{fig:phiRaa_npt} shows the \raa for the $\phi$ meson as a 
function of \Npart in \auau collisions measured over the \pt range $2.5 
< p_T < 5.0$ GeV/$c$ and in the rapidity region $1.2<|y|<2.2$. The 
\raa exhibits an enhancement around intermediate \Npart$\approx100$, while 
remaining largely unchanged elsewhere. These results agree within 
uncertainties with the PHENIX measurements at 
midrapidity~\cite{PhysRevC.83.024909} for both \auau and \cuau 
collisions. While the midrapidity data hints at a possible enhancement 
at lower \Npart, $\approx50$, which is likely due to the lower \pt range 
covered ($2<p_T<5$ GeV/$c$). Nevertheless, this enhancement remains 
consistent with unity when considering the uncertainties. Additionally, 
Fig.~\ref{fig:phiRaa_npt} shows the ALICE \raa of the $\phi$ in \pbpb 
collisions at 2.76 TeV over a similar \pt range, $2<p_T<5$ GeV/$c$, but 
at more forward rapidity, $2.5<y<4$~\cite{Acharya2018}. Although both 
this and the ALICE measurements have considerable 
uncertainties, this result clearly shows an enhancement at intermediate 
\Npart, while the ALICE measurement shows no enhancement. This could 
suggest an energy dependence of strangeness enhancement at forward 
rapidity.

%%%%%%%%%%%%%%%%%%%%%%%%%%%%%%%%%%%%%%%%%%%%%%%%%%%%%%%%%%%%%% Fig_11
\begin{figure*}[htb]
\begin{minipage}{0.48\linewidth}
    \includegraphics[width=1.0\linewidth]{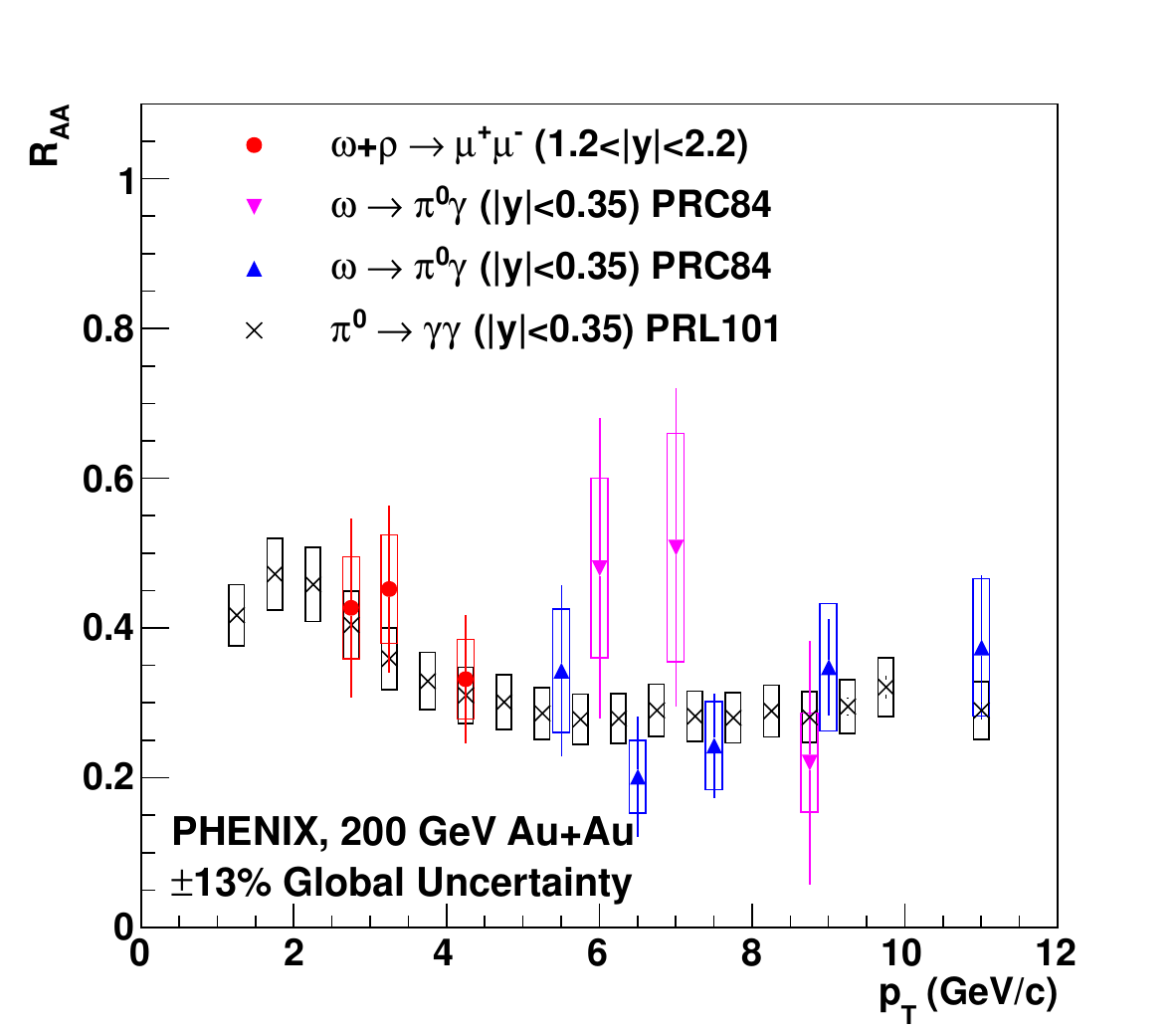}
    \caption{\raa of \omgRho mesons as a function of \pt at 
$1.2<|y|<2.2$ in \auau collisions at \sqsntwo (red points), compared 
with PHENIX $\omega$ measurements at $|y|<0.35$ (blue and inverted 
magenta triangles)~\protect\cite{PhysRevC.84.044902} and $\pi^0$ (black 
stars)~\protect\cite{PhysRevLett.101.232301}.
}
\label{fig:omgRaa_pt}%
\end{minipage}
%\end{figure}
%%%%%%%%%%%%%%%%%%%%%%%%%%%%%%%%%%%%%%%%%%%%%%%%%%%%%%%%%%%%%% Fig_12
%\begin{figure}[htbp]
\hspace{0.3cm}
\begin{minipage}{0.48\linewidth}
    \includegraphics[width=1.0\linewidth]{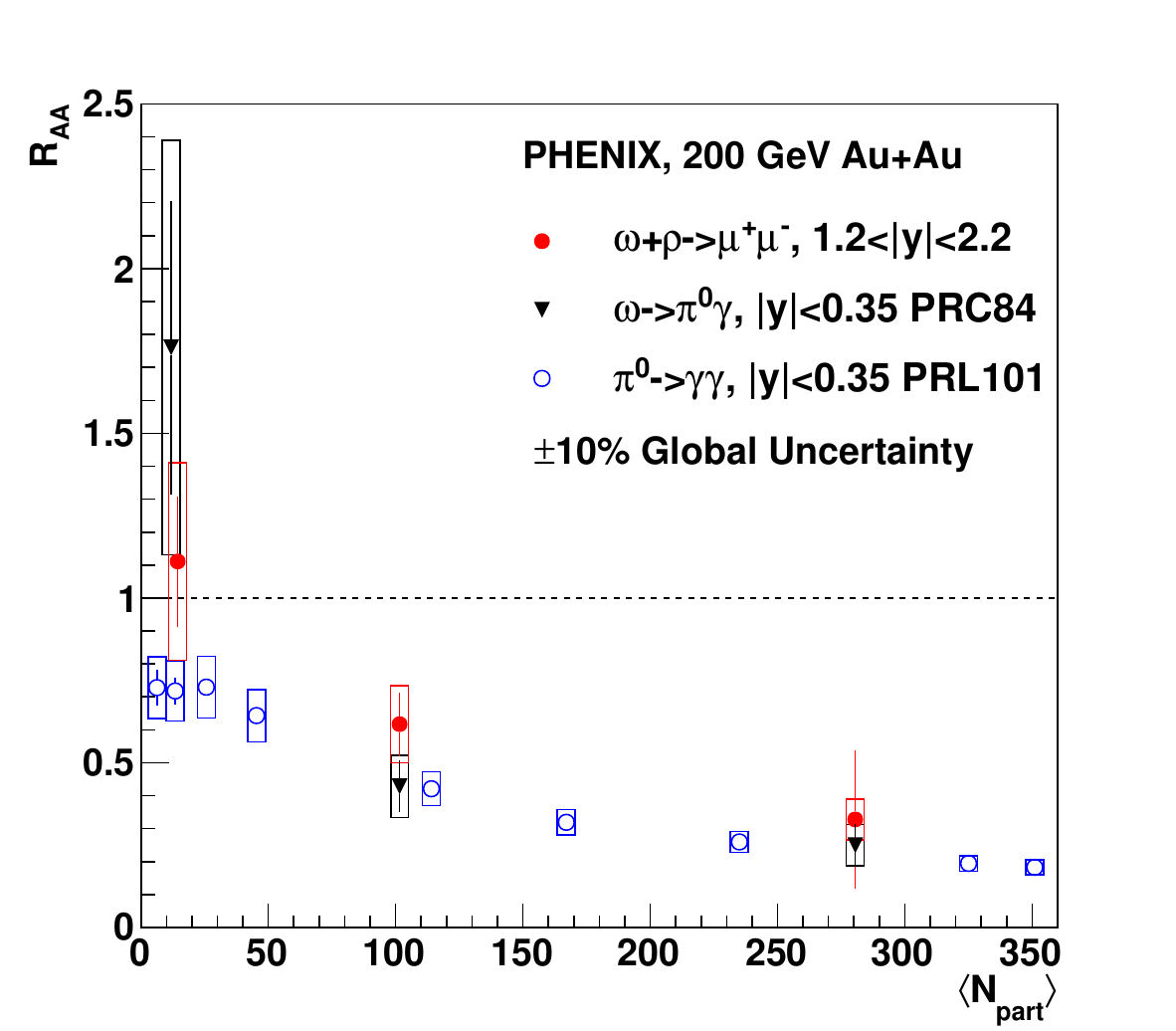}
    \caption{\raa of \omgRho mesons as a function of \Npart in \auau 
collisions, obtained from the dimuon decay channel (solid red points), 
is compared with \raa of $\omega$ at midrapidity obtained from the 
$\pi^0\gamma$ decay channel (inverted black 
triangles)~\cite{PhysRevC.84.044902}. Additionally, the results for \raa 
of $\pi^0$ are shown (empty blue circles) for 
comparison~\cite{PhysRevLett.101.232301}.}%
\label{fig:omgRaa_npt}%
\end{minipage}
%\end{figure}
%%%%%%%%%%%%%%%%%%%%%%%%%%%%%%%%%%%%%%%%%%%%%%%%%%%%%%%%%%%%%% Fig_13
%\begin{figure}[htbp]
\begin{minipage}{0.48\linewidth}
    \includegraphics[width=1.0\linewidth]{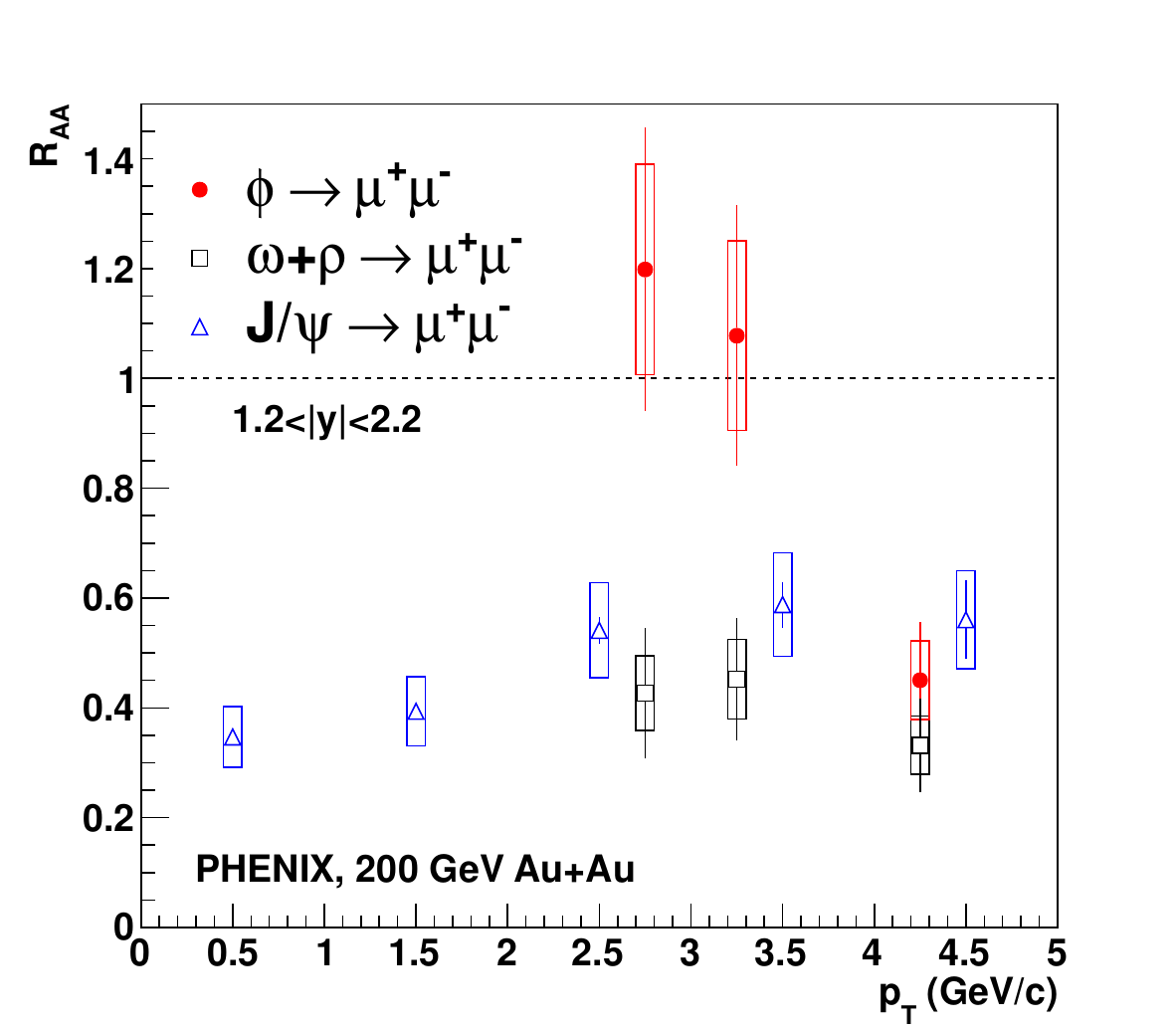}
    \caption{Comparison of \raa as a function of \pt for $\phi$ (solid 
[red] circles), \omgRho (empty [black] squares), and \Jpsi (empty [blue] 
triangles) in mesons \auau collisions at 
\sqsntwo~\protect\cite{PhysRevLett.98.232301}.}
\label{fig:vm_pt}
\end{minipage}
%\end{figure}
%%%%%%%%%%%%%%%%%%%%%%%%%%%%%%%%%%%%%%%%%%%%%%%%%%%%%%%%%%%%%% Fig_14
%\begin{figure}[htbp]
\hspace{0.3cm}
\begin{minipage}{0.48\linewidth}
    \includegraphics[width=1.0\linewidth]{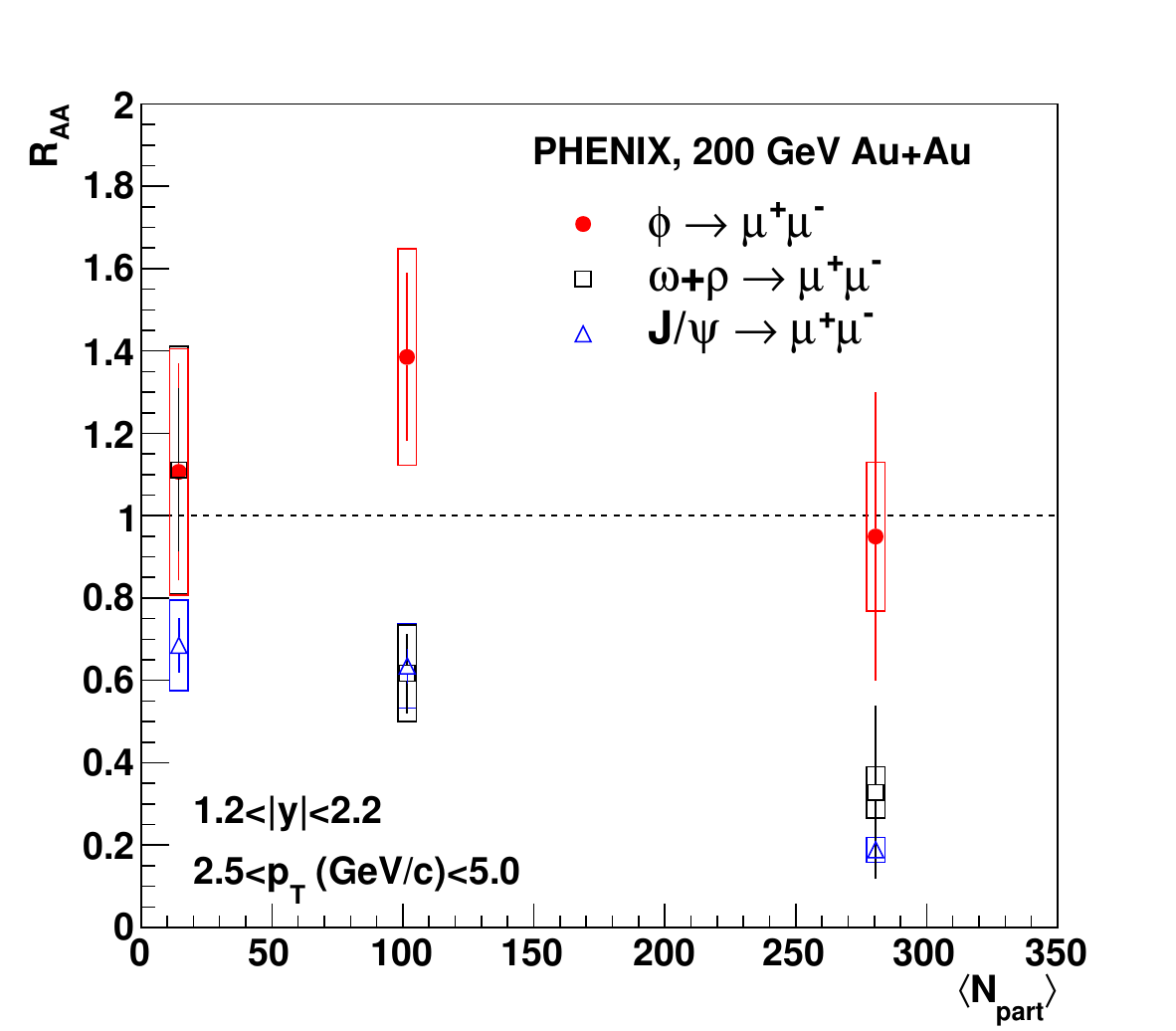}
    \caption{Comparison of \raa as a function of \Npart for $\phi$ 
(solid [red] circles), \omgRho (open [black] squares), and \Jpsi mesons
(open [blue] triangles) in \auau collisions at 
\sqsntwo~\cite{PhysRevLett.98.232301}.}
\label{fig:vm_npt}%
\end{minipage}
\end{figure*}

The \raa of the \omgRho mesons in \auau collisions summed for all 
centralities as a function of \pt, measured over the rapidity range 
$1.2<|y|<2.2$, is shown in Fig.~\ref{fig:omgRaa_pt}, which also shows the 
$\omega$ and $\pi^0$ mesons \raa at 
midrapidity~\cite{PhysRevC.84.044902,PhysRevLett.101.232301}. The \omgRho 
mesons show a similar suppression to that of the $\omega$ meson at 
midrapidity. It should be noted that this measurement, which measures 
the sum of the $\omega$ and $\rho$ mesons, is consistent with the 
$\omega$ meson only, suggesting that the $\omega$ and $\rho$ mesons 
exhibit similar suppression patterns.

The \raa of the \omgRho mesons as a function of the average number of 
participants is shown in Fig.~\ref{fig:omgRaa_npt}. In peripheral 
collisions, \raa is consistent with unity within uncertainties, 
indicating no nuclear modification. In the most central and intermediate 
collisions, \raa is reduced to approximately $0.3$ and $0.6$, 
respectively, showing strong suppression in the most central collisions 
and moderate suppression in intermediate collisions, relative to the \pp 
reference.

Although the PHENIX midrapidity measurement~\cite{PhysRevC.84.044902} 
extends \pt coverage down to 0.3 GeV/$c$, the results remain consistent 
with the current measurement. Figure~\ref{fig:omgRaa_npt} also includes 
the \raa for the $\pi^0$ meson at 
midrapidity~\cite{PhysRevC.84.044902,PhysRevLett.101.232301}. Although 
the $\pi^0$ meson \raa exhibits a suppression pattern similar to that of 
the \omgRho mesons in intermediate and central collisions, the 
suppression persists even in the most peripheral collisions.

Figure~\ref{fig:vm_pt} presents the \raa of the \omgRho, $\phi$, and 
\Jpsi mesons as a function of \pt in the rapidity regions $1.2 < |y| < 
2.2$ in \auau collisions at \sqsntwo~\cite{PhysRevLett.98.232301}. This 
comparison allows for the investigation of flavor dependence in medium 
effects within systems containing light, strange, and charm quarks, 
specifically in the \omgRho, $\phi$, and \Jpsi mesons. The \omgRho 
mesons, representing the light flavor, exhibit the most pronounced 
suppression over the covered \pt range. In contrast, the $\phi$ meson, 
representing the strange flavor, displays a more complex pattern. It 
shows no suppression, maybe a hint of enhancement, in the intermediate 
\pt range, followed by a suppression at higher \pt values that is 
comparable to light flavor. The \Jpsi meson, representing the heavy 
flavor, demonstrates suppression across the full covered \pt range, 
extending even to very low \pt values. However, its suppression pattern 
differs from that of the light flavor. The \Jpsi suppression is more 
pronounced at low \pt and becomes less severe as \pt increases.

A comparison similar to that shown in Fig.~\ref{fig:vm_pt} is presented 
in Fig.~\ref{fig:vm_npt}, but this time as a function of \Npart. In the 
most central collisions, the \omgRho and \Jpsi mesons exhibit 
suppression, while the $\phi$ meson remains largely unchanged. At 
intermediate centralities (around \Npart $\approx 100$), \omgRho and \Jpsi 
mesons similar moderate suppression, while the $\phi$ meson have large 
\raa. In peripheral collisions, \omgRho mesons deviate from the trend 
observed for the \Jpsi meson and show no nuclear modification, behaving 
similarly to the $\phi$ meson in this region.

%%%%%%%%%%%%%%%%%%%%%%%%%%%%%%%%%%%%%%%%%%%%%%%%%%%%%%%%%%%%%%%%%%%%%%%%%%%%%%

\section{SUMMARY}

We have measured the production of \omgRho and $\phi$ mesons at forward 
and backward rapidities via the $\mu^+\mu^-$ decay channel in \pp and 
\auau collisions at \sqsntwo. This study represents the first 
measurement of the invariant yields and nuclear-modification factors for 
\omgRho and $\phi$ mesons in \auau collisions at RHIC in the 
forward/backward rapidity regions. The results are presented as a 
function of \Npart and \pt.

In \auau collisions, the \raa for the $\phi$ meson shows strong 
suppression at higher \pt, while remaining consistent with unity at low 
to intermediate \pt within uncertainties. This behavior is consistent 
with previous findings from \cuau collisions~\cite{PhysRevC.93.024904}. 
Furthermore, the \raa of the $\phi$ meson as a function of \Npart 
reveals distinct trends. An enhancement is observed at intermediate 
\Npart, while it remains largely unchanged at other values. These 
observations are consistent with PHENIX measurements at midrapidity 
within experimental uncertainties. Additionally, measurements by ALICE 
at the LHC~\cite{Acharya2018} show no enhancement across all 
centralities.

In the measured region, the \omgRho mesons exhibits a suppression 
pattern similar to that of the $\omega$ meson as a function of \pt and 
\Npart. In addition, within uncertainties, these results are also 
consistent with those observed for the $\pi^0$ meson. However, it is 
worth noting that the $\pi^0$ meson consistently shows a more pronounced 
suppression. The observed similarity in suppression between \omgRho and 
$\omega$ mesons suggests that the $\rho$ meson may exhibit a similar 
suppression pattern to the $\omega$ meson.

The flavor dependence of medium effects is examined by comparing the 
\raa of the $\phi$, \omgRho, and \Jpsi mesons. Both \omgRho and \Jpsi 
mesons exhibit suppression across the entire \pt range, though their \pt 
dependence differs slightly. Their behavior as a function of \Npart is 
generally similar, except in peripheral collisions where they diverge.  
In contrast, the strange flavor ($\phi$ meson) shows a distinct pattern, 
with enhancement observed at intermediate values of both \pt and \Npart.
 
In conclusion, the analysis provides valuable information on the 
production and nuclear modification of \omgRho and $\phi$ mesons in 
heavy-ion collisions, with both mesons showing significant nuclear 
medium effects, especially at higher \pt and central collisions. These 
results on low-mass vector meson production in relativistic heavy-ion 
collisions contribute to the understanding of particle production and 
the behavior of nuclear matter in extreme conditions.

%\section*{DATA AVAILABILITY}

% The data that support the findings of this article are not publicly available.
% The values in the plots and tables associated with this article are stored in
% HEPData~\cite{hepdata}.

%%%%%%%%%%%%%%%%%%%%%%  ACKNOWLEDGMENTS}  %%%%% MGS25b version
%% 2018 change in Korea
%%% 2021 change in dropping Brazil, Germany, and Pakistan, because
%%%      they no longer have active MGS and left PHENIX before 2015
%% 2024 add HUN-REN ATOMKI [and remove some] (Hungary)
%% added Zambia for PPG258

\begin{acknowledgments}

We thank the staff of the Collider-Accelerator and Physics
Departments at Brookhaven National Laboratory and the staff of
the other PHENIX participating institutions for their vital
contributions.  
We acknowledge support from the Office of Nuclear Physics in the
Office of Science of the Department of Energy,
the National Science Foundation,
Abilene Christian University Research Council,
Research Foundation of SUNY, and
Dean of the College of Arts and Sciences, Vanderbilt University
(U.S.A),
Ministry of Education, Culture, Sports, Science, and Technology
and the Japan Society for the Promotion of Science (Japan),
Conselho Nacional de Desenvolvimento Cient\'{\i}fico e
Tecnol{\'o}gico and Funda\c c{\~a}o de Amparo {\`a} Pesquisa do
Estado de S{\~a}o Paulo (Brazil),
Natural Science Foundation of China (People's Republic of China),
Croatian Science Foundation and
Ministry of Science and Education (Croatia),
Ministry of Education, Youth and Sports (Czech Republic),
Centre National de la Recherche Scientifique, Commissariat
{\`a} l'{\'E}nergie Atomique, and Institut National de Physique
Nucl{\'e}aire et de Physique des Particules (France),
Bundesministerium f\"ur Bildung und Forschung, Deutscher
Akademischer Austausch Dienst, and Alexander von Humboldt Stiftung (Germany),
J. Bolyai Research Scholarship, EFOP, HUN-REN ATOMKI, NKFIH,
MATE KKF, and OTKA (Hungary), 
Department of Atomic Energy and Department of Science and Technology (India),
Israel Science Foundation (Israel),
Basic Science Research and SRC(CENuM) Programs through NRF
funded by the Ministry of Education and the Ministry of
Science and ICT (Korea).
Ministry of Education and Science, Russian Academy of Sciences,
Federal Agency of Atomic Energy (Russia),
VR and Wallenberg Foundation (Sweden),
University of Zambia, the Government of the Republic of Zambia (Zambia),
the U.S. Civilian Research and Development Foundation for the
Independent States of the Former Soviet Union,
the Hungarian American Enterprise Scholarship Fund,
the US-Hungarian Fulbright Foundation,
and the US-Israel Binational Science Foundation.

\end{acknowledgments}

%\bibliography{ppg258x0}

%apsrev4-2.bst 2019-01-14 (MD) hand-edited version of apsrev4-1.bst
%Control: key (0)
%Control: author (8) initials jnrlst
%Control: editor formatted (1) identically to author
%Control: production of article title (0) allowed
%Control: page (0) single
%Control: year (1) truncated
%Control: production of eprint (0) enabled
%
 
\end{document}